\documentclass[manuscript]{acmart}
\usepackage{enumitem}
\usepackage{tikz}
\usepackage{array}
\usepackage{wrapfig}         
\usepackage{newunicodechar}
\usepackage{newunicodechar}
\usepackage{textgreek}

\newunicodechar{β}{$\beta$}

\usepackage[export]{adjustbox}
\usetikzlibrary{external, shapes.geometric, arrows}
\usepackage{xspace}
\usepackage{wrapfig} 

\usepackage{listings}
\usepackage{placeins} % for \FloatBarrier
\usepackage{float}    % for [H] if you 

\usepackage{xcolor}
\usepackage{soul}         
\definecolor{quotegrey}{RGB}{242,242,242}
\sethlcolor{quotegrey}
\newcommand{\pquote}[1]{\hl{\textit{"#1"}}}

\usepackage{tabularx,booktabs,ragged2e,makecell}
\newcolumntype{Y}{>{\RaggedRight\arraybackslash}X}
\newcolumntype{P}[1]{>{\RaggedRight\arraybackslash}p{#1}}

\providecommand{\imgrelpath}{} % e.g., \renewcommand{\imgrelpath}{sectionA/}

\newlength{\miniimgheight}
\newcommand{\usecell}[3]{% #1 mean, #2 sd, #3 filename
  \begin{tabular}[t]{@{}l@{\hspace{0.35em}}c@{}}
    \textbf{#1} (SD=#2) &
    \IfFileExists{parts/figure/\imgrelpath#3}{%
      \includegraphics[height=\miniimgheight,keepaspectratio]{parts/figure/\imgrelpath#3}%
    }{%
      \IfFileExists{parts/figure/#3}{%
        \includegraphics[height=\miniimgheight,keepaspectratio]{parts/figure/#3}%
      }{%
        \rule{2.2cm}{0.85\miniimgheight}% fallback box (keeps row height stable)
      }}%
  \end{tabular}%
}

\newcommand{\useimage}[3]{%
  \raisebox{-0.8\height}{%
    \IfFileExists{parts/figure/\imgrelpath#3}{%
      \includegraphics[width=0.9\linewidth,keepaspectratio,valign=c]{parts/figure/\imgrelpath#3}%
    }{%
      \IfFileExists{parts/figure/#3}{%
        \includegraphics[width=0.9\linewidth,keepaspectratio,valign=c]{parts/figure/#3}%
      }{%
        \rule{3.0cm}{1.2\miniimgheight}% fallback box (aligned)
      }%
    }%
  }%
}

\AtBeginDocument{%
  \providecommand\BibTeX{{%
    \normalfont B\kern-0.5em{\scshape i\kern-0.25em b}\kern-0.8em\TeX}}}

\begin{document}

\tikzstyle{5_box_node} = [
    rectangle,
    rounded corners, 
    minimum width=2cm, 
    minimum height=1cm,
    text centered,
    text width = 2.5cm,
    draw=black,
]
\tikzstyle{3_box_node} = [
    rectangle,
    rounded corners, 
    minimum width=3cm, 
    minimum height=1cm,
    text centered,
    text width = 4cm,
    draw=black,
]
\tikzstyle{4_box_node} = [
    rectangle,
    rounded corners, 
    minimum width=3cm, 
    minimum height=1cm,
    text centered,
    text width = 3.2cm,
    draw=black,
]
\tikzstyle{arrow} = [thick,->,>=stealth]


\newcommand{\yq}[1]{\textcolor{violet}{#1}}

\newcommand{\vm}[1]{\textcolor{blue}{#1}}
\newcommand{\sssec}[1]{\vspace*{0.05in}\noindent\textbf{#1}}

\newcommand{\ToDo}[1]{\textcolor{red}{#1}}
\newcommand{\sys}{\text{FlexMind}\xspace}
\newcommand{\tradeoffmitigation}{\text{Trade-off} $\rightarrow$ \text{Mitigation}\xspace}
\newcommand{\tom}{\text{TO-M}\xspace}

\title{FlexMind: Supporting Deeper Creative Thinking with LLMs}

\author{Yaqing Yang}
\affiliation{%
  \institution{Carnegie Mellon University}
  \city{Pittsburgh, PA}
  \country{USA}}
\email{yaqingyy@cs.cmu.edu}

\author{Vikram Mohanty}
\affiliation{%
  \institution{Carnegie Mellon University}
  \city{Pittsburgh, PA}
  \country{USA}}
\email{vikrammohanty@acm.org}

\author{Yan-Ying Chen}
\affiliation{%
  \institution{Toyota Research Institute}
  \city{Los Altos, CA}
  \country{USA}}
\email{yan-ying.chen@tri.global}

\author{Matthew K. Hong}
\affiliation{%
  \institution{Toyota Research Institute}
  \city{Los Altos, CA}
  \country{USA}}
\email{matt.hong@tri.global}

\author{Nikolas Martelaro}
\affiliation{%
  \institution{Carnegie Mellon University}
  \city{Pittsburgh, PA}
  \country{USA}}
\email{nikmart@cmu.edu}

\author{Aniket Kittur}
\affiliation{%
  \institution{Carnegie Mellon University}
  \city{Pittsburgh, PA}
  \country{USA}}
\email{nkittur@cs.cmu.edu}

\renewcommand{\shortauthors}{Yang et al.}

\begin{abstract}
Expert ideation requires more than generating ideas. People must articulate trade-offs, develop mitigations, and compare parallel directions, yet this  reasoning process is cognitively demanding. 
Existing ideation tools fall short by either only focusing on expanding or organizing the idea space, or by offering evaluations without supporting mitigation generation and continued reasoning over newly introduced trade-offs.
We present FlexMind, an AI-augmented system that proactively scaffolds multi-threaded \tradeoffmitigation chains within a structured workspace. FlexMind helps users surface concrete trade-offs, generate actionable mitigations, navigate and expand multiple idea branches. In a controlled study (n=20), FlexMind supported deeper exploration without sacrificing breadth and produced higher-quality ideas as judged by domain experts, compared to a baseline. A case study with (n=6) experts further reveals how FlexMind supports idea analysis while helping uncover new solutions. These results further inform implications for enhancing human–AI co-creation.

\end{abstract}

\begin{CCSXML}
\end{CCSXML}



\keywords{Human-AI Collaboration, Generative AI, Creativity Support}

\begin{CCSXML}
<ccs2012>
   <concept>
       <concept_id>10003120.10003121.10003129</concept_id>
       <concept_desc>Human-centered computing~Interactive systems and tools</concept_desc>
       <concept_significance>500</concept_significance>
       </concept>
   <concept>
       <concept_id>10003120.10003121.10003122</concept_id>
       <concept_desc>Human-centered computing~HCI design and evaluation methods</concept_desc>
       <concept_significance>500</concept_significance>
       </concept>
 </ccs2012>
\end{CCSXML}

\ccsdesc[500]{Human-centered computing~Interactive systems and tools}
\ccsdesc[500]{Human-centered computing~HCI design and evaluation methods}

\begin{teaserfigure}
  \includegraphics[width=\linewidth]{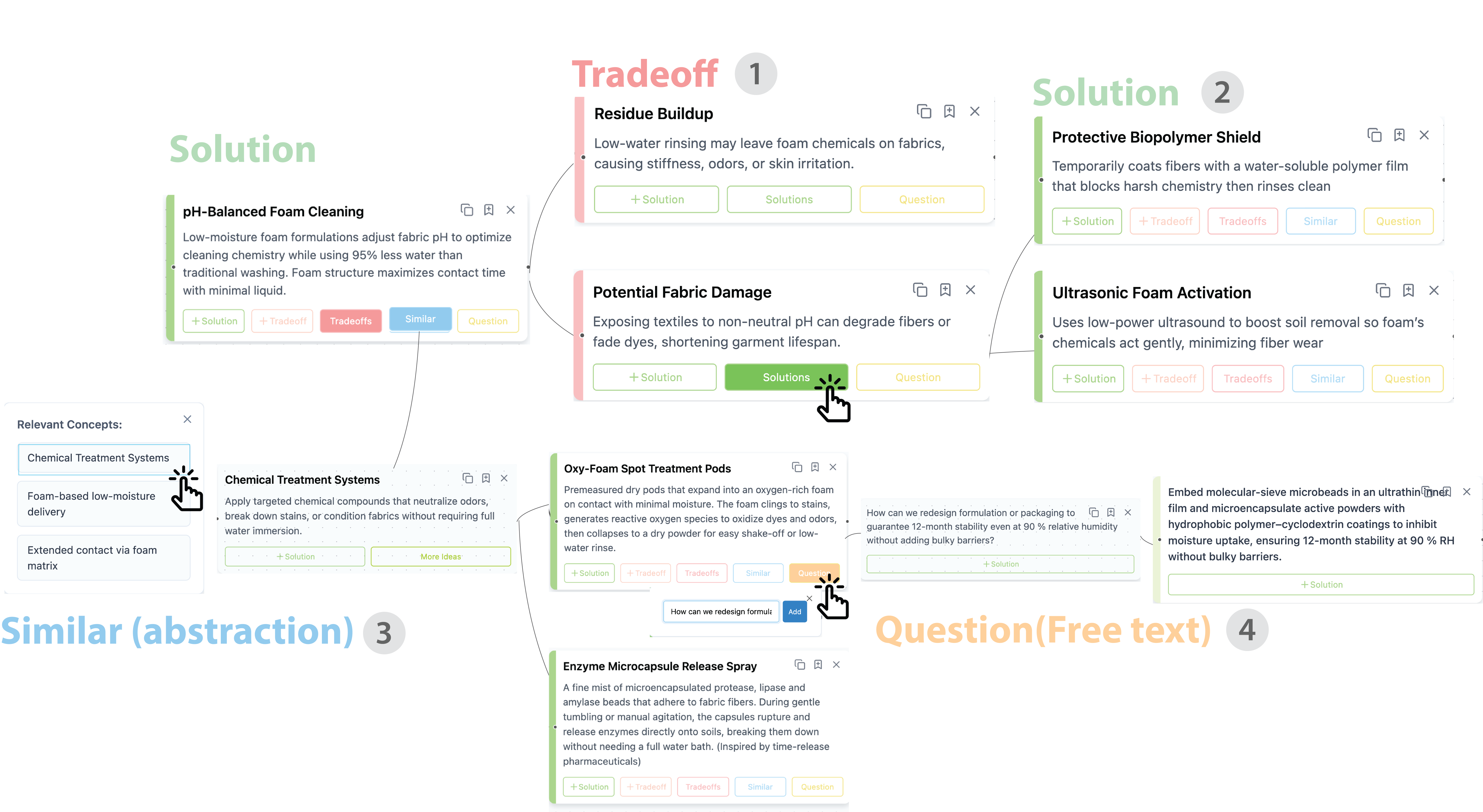}
  \caption{\textbf{\sys Overview.} The system supports deep and broad exploration of the solution space during ideation through a multi-threaded \tradeoffmitigation structure. Four core affordances scaffold the structure: (1) \textbf{Trade-off analysis} expands a solution into potential risks (red cards); (2) \textbf{Mitigation suggestions} extend a trade-off into candidate solutions (green cards); (3) \textbf{Similar-by-schema} abstracts the current idea to seed alternative branches; and (4) \textbf{Card-level Q\&A} supports free-text queries stored on the originating card. Green denotes solutions and red denotes trade-offs.}
  \Description{A canvas view with clustered “idea cards.” Green cards show solutions; red cards show trade-offs. Callout 1 highlights a solution with nearby red trade-off cards. Callout 2 highlights a red trade-off with new green mitigation cards fanning out. Callout 3 points to a “Relevant Concepts” panel listing high-level categories used to find similar ideas. Callout 4 points to a small text box for asking a free-form question whose answer attaches to that card. The layout depicts how schemas give breadth while a tree of linked cards captures depth.}
    \label{fig:overview}
\end{teaserfigure}
  % \caption{\sys supports ideation in breadth and depth: (1)organizes ideas into schemas and captures actions in a tree structure for tracking and engaging in multiple thinking threads;(2)leverages tradeoff–solution as the core structure of information chains to enable evaluation and elaboration by (a) suggesting tradeoffs and mitigation strategies, (b) abstracting ideas to surface diverse alternatives, and (c) supporting customized queries for instant information.}

\maketitle

\label{intro} 

\section{Introduction}
Creative breakthroughs rarely emerge fully-formed; instead, they require rigorous effort to navigate inherent challenges and tradeoffs. James Dyson, for example, found his initial inspiration in the industrial-scale cyclones used in sawmills. However, translating that "spark" into a viable consumer product required managing a series of trade-offs: solving the challenge of scale via a single tangential inlet, for instance, introduced new problems such as diminished separation efficiency and increased airflow resistance. Rather than fixating on a single path, Dyson explored over 5,000 different prototypes across multiple idea threads, eventually arriving at a successful design~\cite{dyson2021invention}.

This pattern reflects a broader reality of expert ideation. Early concepts often function as preinventive structures—initial frameworks that possess inherent flaws or functional trade-offs~\cite{finke1996creative, ward1999creative}. Addressing one limitation frequently reveals new technical issues, creating a "chain" of iterative refinements~\cite{tromp2024creativity}. However, success rarely stems from deep-diving into a single initial thought; instead, experts must balance ideation depth with ideation breadth~\cite{dow2010parallel, brophy2001comparing} to both unfold a wide array of possibilities and work through their initial limitations to find the sweet spots of the latent design space ~\cite{de1970lateral}.

This process presents several challenges. First, identifying the trade-offs inherent in early ideas is difficult~\cite{amabile1983social,mumford2003have}. Many design tasks are interdisciplinary and complex, so even domain experts may encounter aspects of a problem where their knowledge is limited~\cite{cross2004expertise}, making flaws hard to recognize. This difficulty is compounded for ideas designers generate themselves, as confirmation bias~\cite{hallihan2012confirmation} and emotional attachment~\cite{liedtka2015perspective} to promising aspects of an idea can lead them to overlook or avoid considering potential weaknesses. Even when designers do become aware of such flaws, they may abandon an idea chain prematurely due to limited domain knowledge that obscures possible mitigations~\cite{cross2004expertise}, cognitive overload when reasoning through chains in depth~\cite{leschziner2019beyond} or time constraints. At the same time, limited working memory~\cite{cowan2012working} can restrict breadth of exploration, making it challenging to consider multiple idea threads in parallel.

A rich history of ideation support systems have attempted to address various components of the challenges above. A large number of tools have explored the design space of generating expansive idea collections~\cite{di2022idea, hou2024c2ideas, lawton2023drawing} or organizing solutions into high-level categories~\cite{suh2024luminate, siangliulue2016ideahound}. These tools have demonstrated the value of visualizing multiple ideas at once, allowing users to externalize their mental models, switch between idea threads, and reduce working memory load. %Critically, they often overlook the support for the iterative process of identifying trade-offs and generating corresponding mitigations.
% Approaches to support identifying flaws in ideas have begun to appear, one line of work develops tools that use pre-specified rubrics to score ideas along dimensions such as feasibility and value~\cite{shaer2024ai,collier2025good}.Others use large language models to more flexibly surface trade-offs and potential issues for a given idea on demand~\cite{kang2024biospark,collier2025good}. 
A related line of work goes beyond visualizing or clustering ideas to structuring them into links or chains, such as linking ideas to each other~\cite{shen2025ideationweb} or supporting various link types like subproblem-solution pairs~\cite{chen2025coexploreds}. This work lays a conceptual foundation for computationally supporting structured threads in ideation. While these and other approaches have explored a variety of structure and link types, we are unaware of work that has specifically investigated computational support for the structures we explore here: proactively generating the recursive, parallel chains of trade-offs and mitigations as commonly seen in expert design. Recently, approaches have begun to emerge that enable some of the necessary functions of creating such chains, such as identifying the flaws or tradeoffs in a given idea using pre-specified rubrics that score feasibility and value~\cite{shaer2024ai, collier2025good} or using large language models (LLMs) to identify both tradeoffs and potential mitigations on demand~\cite{kang2024biospark, collier2025good}.
However, these approaches treat tradeoffs as one-offs, providing independent scores or issues and requiring user initiative to elicit and work through them. In this work we explore the idea of going beyond the first link  to proactively scaffold chains of further depth, and support the flexible exploration of many such chains, leveraging the LLMs.

In doing so we acknowledge the potential double-edged nature of using LLMs for creativity support. While our work as well as many of the ideation tools cited above leverage the generative power of Large Language Models (LLMs)~\cite{shen2025ideationweb,liu2024toward,chen2025coexploreds,collier2025good,kang2024biospark}, there is growing evidence showing that using LLMs could potentially degrade the critical thinking and creativity~\cite{wadinambiarachchi2024effects,lee2025impact,kumar2025human,tankelevitch2025understanding} 
% \textbf{[CITE papers here, including the now classic microsoft paper i think it is Lee et al and advait sarkar is on it, as well as our workshop proceedings, also the work on homogeneity]}. 
Interacting with LLMs can lead to users overly accepting generated ideas without further evaluating or exploring them~\cite{kim2023effect} or developing an over-reliance on automated outputs \cite{li2024user, kumar2025human}. LLM-generated output can also evidence a surface-level understanding of context, as well as suffer from homogeneity effects ~\cite{wadinambiarachchi2024effects}, potentially limiting its value to experts who can tell when it is wrong and potentially misleading novices who cannot. Our work aims to mitigate these issues by exploring how user judgment might be integrated into the reasoning process by tightly coupled user-AI interactions leveraging the \tradeoffmitigation structure, making the evaluation and evolution of ideas explicit, and keeping users engaged in active reasoning rather than passive acceptance. 

Another goal of our work here is to advance our knowledge of how AI might help with ideation support in more expert, professional contexts. Although a longstanding issue in creativity support research, given the recent speed of advances in AI models, it is perhaps unsurprising that most AI ideation systems have been evaluated only with relatively novice or student participants, and outcomes are often assessed using proxy measures such as AI judges~\cite{fukumura2025can} or novice user self-evaluation~\cite{shen2025ideationweb}. These evaluation practices leave open the important question of whether LLMs can provide value to experts in complex design and R\&D settings. To address this question we recruited participants who were experienced design and engineering professionals, assessed their work with even more experienced domain experts, and include both a controlled study comparing our tools versus a typical LLM baseline as well as in-situ case studies of practitioners using our tool on their own real-world project ideas.

In this paper we present FlexMind, an AI-augmented system that scaffolds multi-threaded \tradeoffmitigation reasoning for supporting exploration depth and breadth in ideation. FlexMind scaffolds the process in three ways. First, it supports exploration depth by proactively identifying trade-offs and their possible mitigations. Second, it facilitates exploration breadth using a tree structure and high-level schemas to organize multi-threaded \tradeoffmitigation chains. This allows users to engage with different threads, generate new threads when needed, and retain an overview of the evolving idea space without overwhelming working memory. Third, it provides mixed-initiative support for users to fluidly incorporate their own ideas and thinking into the reasoning process. This guides the ideation trajectory and provides context for the AI to think alongside the user. 
More broadly, this work advances our understanding of how AI systems for creativity support can move beyond linear co-creation workflows toward structured, exploratory engagements that support critical yet unconstrained creative reasoning. Concretely, our contributions include:
\begin{itemize}
\item \textbf{A system that proactively scaffolds continuous \tradeoffmitigation reasoning during ideation}, intentionally surfacing idea trade-offs, their mitigations, and subsequent trade-offs as structured reasoning chains. This supports ideation depth within individual ideas while also enabling ideation breadth by revealing alternative solution directions implied by different mitigations.
% deeper exploration of individual ideas by encouraging evaluation and refinement, while also revealing alternative solution directions implied by the mitigations of each trade-off.
\item \textbf{An interface supporting exploration along the multi-threaded structure of ideas} that supports users to flexibly switch between drilling down within a single idea and moving across parallel idea threads to compare, shift, and develop multiple alternatives. 
\item \textbf{An expert-centered evaluation} including a controlled study with professionals and expert case studies, demonstrating that \sys improves idea quality (as judged by domain experts) and revealing how it supports expert ideation in real-world design projects.
\end{itemize}

Our controlled study results suggest that \sys supports deeper exploration of ideas than a ChatGPT baseline without sacrificing breadth, leading to higher idea quality. The expert case study provides detailed insights into how \sys supports professional ideation by reducing ideation stalls through surfacing unfamiliar but relevant knowledge, overcoming working memory limits when managing and comparing multiple ideas, and seeding new idea chains through abstraction when experts needed to break fixation or clarify early concepts. Based on these findings, we further discuss the design implications around the core features of \sys and potential future directions.

\section{Related Work and Design Goals}\label{sec:relatedwork}

\subsection{Working Through Trade-offs, Consequences, and Mitigations}
Transforming raw ideas into viable concepts requires rigorous evaluation, yet bottlenecks persist at the "fuzzy front end" of development where designers must assess feasibility with limited information~\cite{khurana1997integrating}. This evaluation is cognitively demanding. Without support, teams prone to confirmation bias may overlook critical flaws or default to familiar options rather than refining novel ones~\cite{rietzschel2010selection,mueller2012bias}. Formal systems engineering methods address this through structured analysis. Trade studies compare alternatives across performance and cost~\cite{kapurch2010nasa,baker2012survey}, while methods like FMEA~\cite{mikulak2017basics} and premortems~\cite{klein2007performing} anticipate failures to plan mitigations. Set-based design or concurrent engineering further refines this by treating feasibility as an iterative loop of trade-off, mitigation, and reassessment~\cite{toche2020set,sobek1999toyota,iso2018iso}. However, these methods demand high cognitive effort and resources, limiting their application during rapid early-stage ideation.

Recent tools leveraging LLMs attempt to support the early idea evaluation by producing quick critiques, ratings, and hazard analyses~\cite{shaer2024ai,collier2025good,liu2025personaflow}. While faster, these tools rarely prescribe concrete, step-by-step mitigations. In risk-sensitive domains, LLM-generated rationales are often generic and not necessarily reliable, requiring significant human oversight to ensure soundness~\cite{collier2025good}. Interactive tools such as CoExploreDS~\cite{chen2025coexploreds} and Supermind Ideator~\cite{heyman2024supermind} structure ideation by linking problems to solutions. While they could surface useful questions, they leave the core cognitive work: interpreting trade-offs and deriving specific mitigations, entirely to the user. A critical gap remains: existing tools lack deliberate support for the rigorous process of working through a specific trade-off to improve a concept. To address the gaps, \textbf{Our first design goal is to lower the cognitive burden of early-stage evaluation by explicitly scaffolding people's reasoning through tradeoffs, and transforming tradeoffs into mitigation paths using a structured \tradeoffmitigation representation.}

\subsection{Exploring Across Idea Threads with Structured Representations for Broad Exploration}
Despite the necessity of deep \tradeoffmitigation reasoning, expert ideation rarely occurs in a single linear thread. It also requires managing multiple such chains in parallel to compare alternatives and explore the solution space broadly. However, existing tools often force a compromise between the depth of reasoning and the breadth of exploration required for innovation.

Tools designed for breadth could present  overviews of ideas but often lack the specific causal logic of trade-offs and mitigations. Systems like \textit{Luminate} generate and organize design spaces into high-level dimensions~\cite{suh2024luminate}, \textit{IdeaHound} clusters contributions to surface thematic groupings~\cite{siangliulue2016ideahound}, and \textit{WordTree} visualizes hierarchical branching~\cite{linsey2012design}. While these systems make multiple threads co-present, the connections they visualize are usually taxonomic, which reveal how ideas are related by different topics, but they do not expose the rigorous chain of trade-off and mitigation needed to evaluate them. Consequently, users can survey broad topics but cannot easily compare the specific feasibility trade-offs of parallel options.

Conversely, tools designed for depth support the logical progression of ideas but often isolate the user within a single thread. Systems like \textit{CoExploreDS} model problem–solution co-evolution~\cite{chen2025coexploreds}, tools based on IBIS framework analyze ideas using pros and cons~\cite{nomura2024towards}, and AI-assisted \textit{Causal Pathway Diagrams} link interventions to consequences~\cite{zhong2024ai}. These tools successfully scaffold the deep reasoning chains described in the previous section. However, they typically treat the chain as a deep dive into one concept rather than a comparative object. This limitation is particularly evident in standard linear interfaces, such as chat-based workflows, where focusing on the depth of one mitigation often buries the context of parallel alternatives~\cite{meincke2025chatgpt,nguyen2025feedstack}. This isolation makes it cognitively difficult to transfer mitigation strategies across distinct threads or to evaluate multiple "viable futures" side-by-side.

A critical gap remains: breadth-oriented maps rarely propagate the specific consequences of critiques, while depth-oriented traces make it difficult to compare and generate parallel alternatives. \textbf{Our second design goal is to bridge this gap by scaffolding multi-threaded exploration based on the \tradeoffmitigation structure from the first design goal. We aim to make parallel chains co-present, enabling designers to broadly compare different idea paths while preserving the clear traces of trade-offs and mitigations needed to deeply refine them or leverage existing ones to generate new directions that expand the idea space.}

\subsection{Designer-Centered Use of LLMs During Ideation}
LLMs can broaden idea exploration by drawing on extensive internal knowledge for large-scale idea generation~\cite{si2024can,zhou2025examining,ko2023large}. 
Furthermore, LLMs have been shown to support idea evaluation \cite{shaer2024ai,gordetzki2025llm}.
% using different argumentation strategies \cite{gordetzki2025llm}.
Although LLMs are increasingly used to support the ideation stage of various problem-solving tasks~\cite{sinlapanuntakul2025impacts}, research has shown that using LLMs without caution can undermine creativity~\cite{kumar2025human}. Even when generating many ideas, outputs often remain homogenized or underdeveloped~\cite{kim2023effect}. A proposed strategy for overcoming the limitations that LLMs have on their own is to have humans work with AI systems, with the goal of leveraging users' critical evaluation of ideas to ultimately produce better outcomes. 
Yet studies show people tend to over-accept AI suggestions and engage only superficially with considering deeper aspects of LLM-generated concepts~\cite{kim2023effect,wadinambiarachchi2024effects}. 
In early design stages, when knowledge about potential design options and the structure of the design space is limited, such reliance can lead to fixation and narrow exploration~\cite{doshi2024generative}.
Over time, leaning on AI for ideation may weaken a designer's independent reasoning~\cite{kosmyna2025your}.

To counter these risks, researchers have explored human–AI co-creation systems that aim to augment, rather than replace, human thinking~\cite{suh2024luminate}. 
Some approaches use AI to encourage reflection on generated ideas~\cite{xu2025productive}, while decision-making research applies cognitive forcing interventions to reduce overreliance and promote active engagement~\cite{buccinca2021trust}.
Many systems move away from linear chat interfaces, toward more spatial representations common among creative design activities \cite{chen2025coexploreds,pu2025ideasynth}.
Such interfaces afford designers with an overview and representation of the design space in contrast to a rolling, convergent window view on a chat-based conversation.
Building on these prior works and to address issues of designer overreliance on LLM "thinking," \textbf{our third design goal is towards a human–AI partnership model: embedding AI into a visual workspace that probes and extends users' reasoning without deciding for them, sustaining engagement and preventing premature convergence.}

\section{\sys}
We build upon the design goals in Section~\ref{sec:relatedwork} to introduce \textbf{\sys, an ideation tool that organizes thinking through multi-threaded \tradeoffmitigation chains}. The system helps designers articulate tradeoffs, turn trade-offs into actionable mitigations, and explore a broad idea space through a visualized multi-thread structure where new branches are generated through abstraction along the chain and people can move flexibly across chains to follow their thoughts without disrupting or losing existing branches. The process is human led and AI supported so users retain control of exploration while AI extends their reasoning in productive and coherent ways. Together, these components help designers survey a wide space of possibilities, drill down on selected ideas, and navigate across branches while maintaining a coherent view of the overall portfolio. We begin with an example walkthrough and then detail how each component supports
breadth, depth, and designer agency through the multi-threaded \tradeoffmitigation structure.

\subsection{Example Usage Scenario}

\begin{figure}[t]
  \centering
  \includegraphics[width=\linewidth]{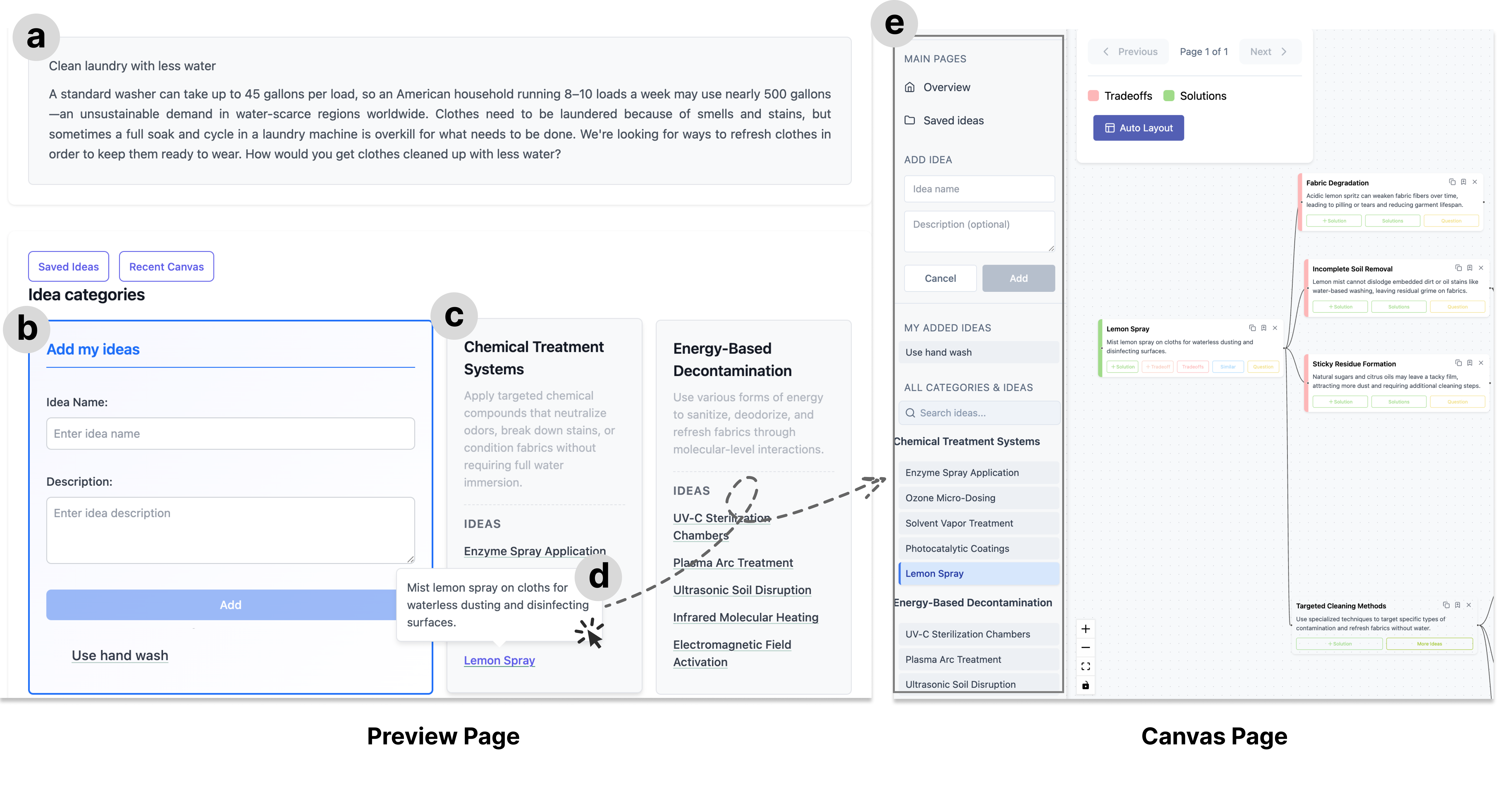}
  \caption{\textbf{\sys: Preview and Canvas Pages.} The preview page (left) presents the design brief (a), allows users to (b) add their own ideas, and (c) browse system-generated ideas organized by schemas (categories). Users can (d) select specific ideas to transfer into the workspace. The canvas page (right) displays selected ideas as linked cards in branchable idea trees on the canvas, while the sidebar (e) organizes categories, saved ideas, and search to support breadth-oriented exploration. Green cards denote solutions and red cards denote trade-offs.}
  \Description{Side-by-side screenshots. Left: Preview page. At the top is the design brief (a). Below, a form lets users add their own ideas (b). To the right, system-generated ideas are grouped by schemas (categories) such as “Chemical Treatment Systems” (c). A pointer shows a user selecting the “Lemon Spray” idea (d). Right: Canvas page. A sidebar (e) lists categories, saved ideas, and a search bar. In the main workspace, ideas appear as cards linked into branchable trees: green cards represent solutions and red cards represent trade-offs. The figure illustrates moving from a schema-level overview to a canvas for in-depth exploration.}
  \label{fig:interface_overview}
\end{figure}

Sarah, an R\&D expert, is tackling the challenge of \emph{cleaning laundry with less water}. She opens \sys and enters a brief description of the design challenge (Figure~\ref{fig:interface_overview}a). \sys generates 50 ideas spread across 10 categories employing different solutions to the problem (e.g., chemical treatment vs. energy-based decontamination; see Figure~\ref{fig:interface_overview}c). Rather than scanning a long list of unorganized and overlapping ideas, using this partitioning of the design space she quickly focuses on threads of interest to decide which ideas merit deeper evaluation.

\begin{figure}[t]
  \centering
  \includegraphics[width=\linewidth]{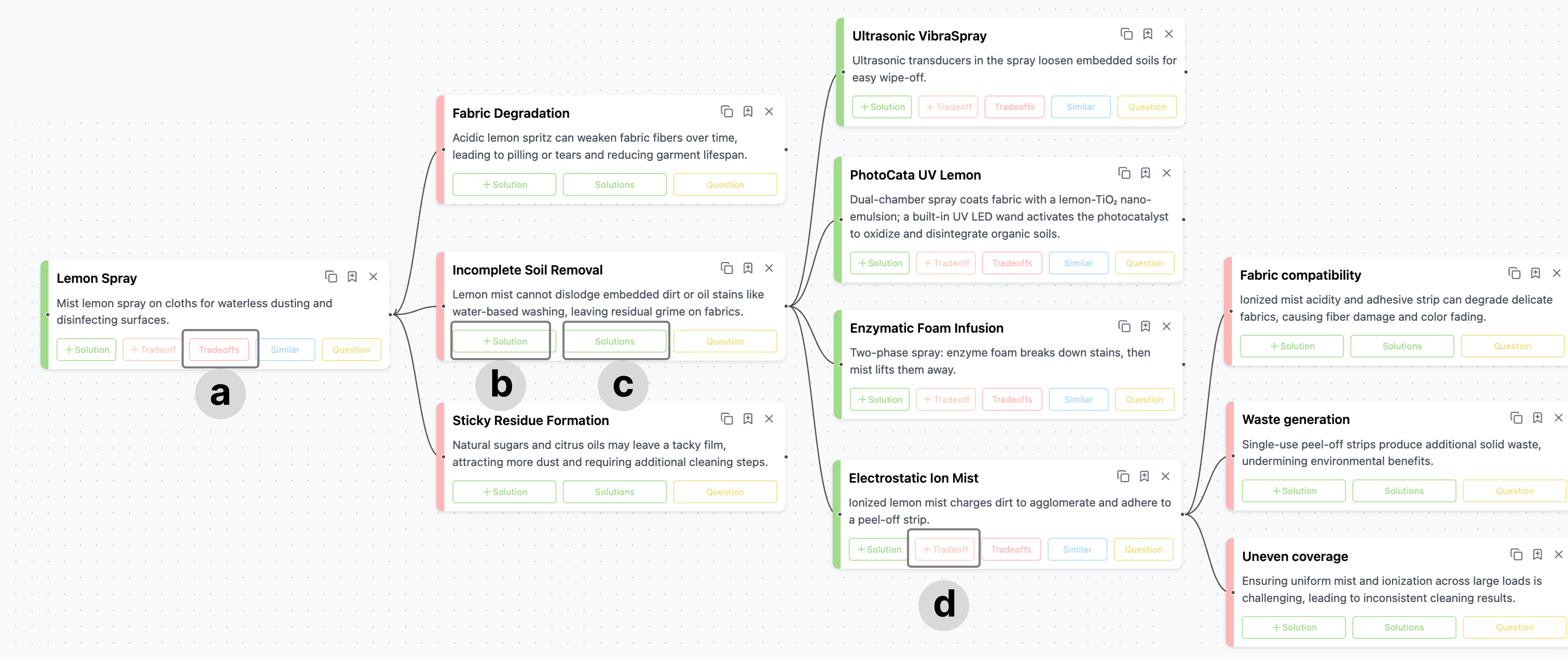}
    \caption{\textbf{Trade-off $\rightarrow$ Mitigation Exploration in \sys.} Users begin a trade-off analysis for a given solution card (e.g., “Lemon Spray”) by clicking the \textbf{(a) Trade-off} button to reveal potential limitations (red cards). To address a trade-off, they can click the \textbf{(c) Solution} button, which generates targeted mitigation ideas (green cards). Users may also \textbf{(b) contribute their own solutions} or \textbf{(d) add new trade-offs}. The system maintains context of the design task while supporting iterative exploration of trade-offs and mitigations as branching idea trees.}
    \Description{A canvas workspace shows an idea tree branching from a green solution card labeled “Lemon Spray” (a). The card’s Trade-off button expands several red trade-off cards, including “Fabric Degradation” and “Incomplete Soil Removal.” From these red cards, users can click the Solution button (c) to generate green mitigation cards, such as “Ultrasonic VibraSpray” and “Enzymatic Foam Infusion.” Users may also add their own solution cards (b) or create new trade-offs (d), shown by the green “Electrostatic Ion Mist” card and its linked red trade-offs. Green cards represent solutions, red cards represent trade-offs, and arrows indicate branching paths of the trade-off → mitigation loop.}
      \label{fig:tradeoff}
\end{figure}

One idea catches her eye: using lemon spray as a low-cost, natural, and sustainable chemical deodorizer. She knows that there are probably hurdles to overcome for practical use, but instead of them causing her to discard the idea immediately she uses \sys will help her think through the idea, so she creates a canvas from it. Clicking the \textbf{Tradeoff} button (Figure~\ref{fig:tradeoff}a) results in the system generating three potential issues to address: \emph{Fabric Degradation}, \emph{Incomplete Soil Removal}, and \emph{Sticky Residue Formation}. Concerned about \emph{Incomplete Soil Removal} she clicks the \textbf{Solution} button on its card (Figure~\ref{fig:tradeoff}c) and the canvas expands with \emph{three green \textbf{solution} cards}, each suggesting a concrete mitigation strategy such as combining it with an ultrasonic transducer or ionizing the lemon mist so that dirt will agglomerate and adhere to a pull-off strip. Considering the idea in more depth helps her see not just whether the idea is viable, but how it might be adapted and improved, transforming trade-off into forward progress.

Curious about other limitations, she runs a trade-off analysis again. The canvas generates three additional red cards: \emph{Fabric Compatibility}, \emph{Waste Generation}, and \emph{Uneven Coverage}, each surfacing a distinct risk that might otherwise be overlooked. Reading through these, Sarah realizes a further concern: \emph{whether repeated lemon exposure could leave lasting odors or discoloration on certain fabrics}. She records this as her own trade-off using \textbf{+Trade-off} (Figure~\ref{fig:tradeoff}d), so her domain knowledge and thought process is tracked alongside the system’s analysis.

\begin{figure}[t]
  \centering
  \includegraphics[width=\linewidth]{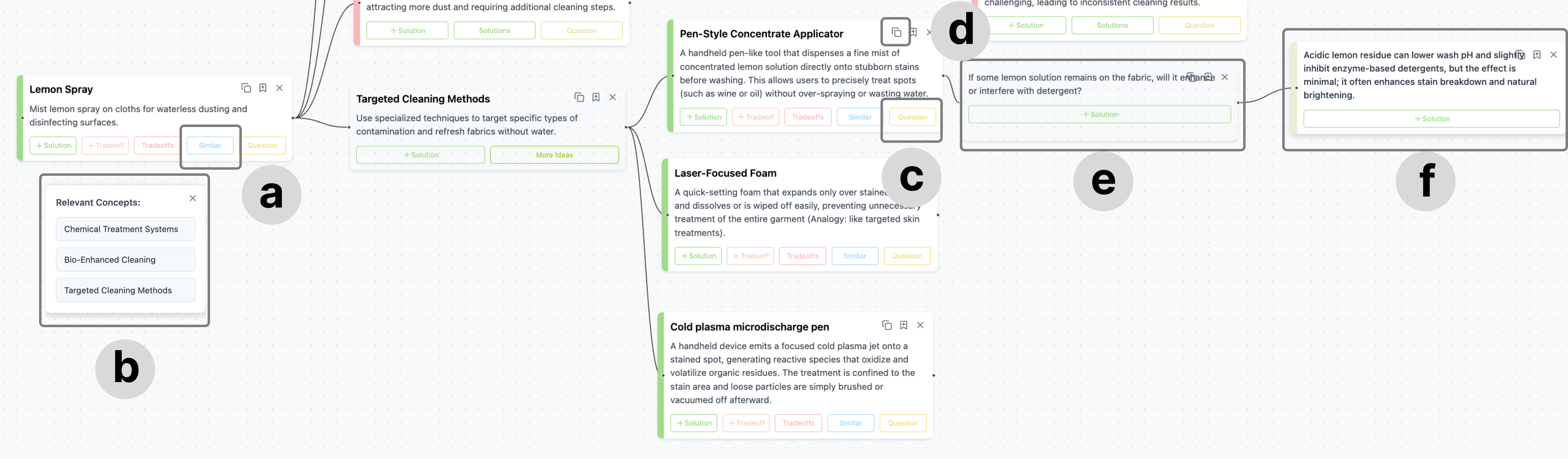}
  \caption{\textbf{Schema-based Exploration and Q\&A in \sys.} Users can surface high-level schemas to find similar ideas: selecting the (a) Similar button displays high-level schemas (b), which expand into alternative solutions. To probe further, users can (e) input free-form queries through the Q\&A feature; with context from the originating card, the system returns  targeted answers (f). Together, schema-level abstraction and card-level Q\&A help users broaden exploration while obtaining instant, context-sensitive information.}
  \Description{A canvas workspace shows how users explore similar ideas and ask questions. On the left, a green “Lemon Spray” solution card is highlighted with the \textbf{(a) Similar} button. Below it, a panel displays \textbf{(b) high-level related categories} such as “Chemical Treatment Systems,” which can expand into additional solution cards. Several related solution cards (c, d) are shown branching outward. On the right, a user enters a free-form query into the \textbf{(e) Q\&A} box, and the system returns a \textbf{(f) targeted answer} attached to the originating card. Green cards represent solutions, and red cards represent trade-offs. The figure illustrates schema-based abstraction for broadening exploration alongside card-level Q\&A for instant, context-sensitive answers.}
  \label{fig:similar}
\end{figure}

Stepping up a level, Sarah realizes that the lemon spray solution actually embodies several different aspects of the design space that might be interesting to explore. Clicking the  \textbf{Similar} button on the original \emph{Lemon Spray} card (Figure~\ref{fig:similar}a) surfaces other active ingredients that lemon spray uses in its solution, such as \emph{Bio-Enhanced Cleaning} (i.e., using ingredients found in nature) and \emph{Targeted Cleaning Methods} (i.e., cleaning only the soiled parts of the clothing). Following the latter path reveals solutions further exploring that part of the design space while opening up others: a \emph{pen-style concentrate applicator}, \emph{laser-focused foam}, and a \emph{cold-plasma microdischarge pen}. While the pen stands out because it could feel familiar to consumers (like a stain-removing pen), she wonders whether any leftover lemon compounds would interact with detergent in a later rinse. She clicks the \textbf{question} button on its idea card, types in her question, and sees the reply attached to that card-that any residue would likely have a \emph{minimal} effect on standard detergents (Figure~\ref{fig:similar}e). This lets her complement her expertise with targeted knowledge specific to the idea at hand, while keeping the rationale visible for later reference. Wanting to switch from deeply exploring of \emph{Lemon Spray} to a wider breadth of solutions, she uses the sidebar to hop to other concepts and pursue different directions. After exploring a range of ideas, she opens the \textbf{Saved Ideas} tab on the side bar (Figure~\ref{fig:interface_overview}e) to scan all her bookmarked ideas, shortlist favorites, and push selected directions into further brainstorming and development.

\subsection{\tradeoffmitigation for Deepening Ideas}
A central step in developing early ideas is recognizing the key tradeoffs they must overcome and considering how those tradeoffs might be addressed~\cite{schulz2018interactive,specking2018early,collier2025good,schon1986reflective}. \sys structures this process through a \tradeoffmitigation chain so that users can surface trade-off while also making progress on next steps.

The \textit{Tradeoff} button on a card triggers an analysis of the selected solution in the context of the design problem. The prompt instructs the LLM to identify all major tradeoffs and then return the top three (see Prompt in Appendix~\ref{prompt_tradeoff_generation}). These results are shown in small batches to keep the information manageable, and repeated clicks generate additional tradeoffs while using previously shown ones as context to avoid redundancy.

Users can then continue the chain by generating mitigations for a selected tradeoff. In the example above, the tradeoff of incomplete soil removal for lemon spray led to mitigations such as using ultrasonic waves to dislodge the dirt, applying a UV activated nano emulsion, and ionizing the dirt to make it easier to remove. When the \textit{Solution} button is clicked, the LLM is prompted to propose solutions grounded in both the tradeoff and the context of the original idea. The prompt guides the model to give solutions that are feasible and clearly explained but not overly detailed so the user can still build on them(see Prompt in Appendix~\ref{prompt_solution_generation}). To reduce fixation, the prompt also encourages both incremental refinements of the current idea and more distinct alternatives. As with tradeoffs, \sys returns three solutions at a time and allows users to request more in controlled batches.

\subsection{Multi-threaded structure for Broadening Ideas}
Building on the \tradeoffmitigation chain that supports deeper reasoning within an idea, \sys helps users explore a broad space of possibilities through a multi-threaded structure. This structure keeps parallel idea paths co-present so users can compare directions and open new branches without disrupting the ones they have already developed. To support the new idea generation, \sys also uses a shared set of high-level categories that serve as a common backbone for generating ideas that start new branches, both in the initial idea set and during later expansions.

The multi-threaded structure begins on the initial page, where users can choose different seed ideas as starting points for exploration. Each seed leads to its own \tradeoffmitigation chain on the canvas, allowing users to map out different directions in parallel. To generate these seeds, \sys takes the user’s brief and runs a three step prompt chain to produce ten broad categories and five concrete ideas within each one (see Prompts in Appendix~\ref{prompt_schema_gen}--\ref{prompt_schema_idea_gen}). The aim is to give an early map of distinct approaches rather than a single entry point. These categories appear in the sidebar and anchor the first set of idea threads.

As users explore, they can also generate new branches by abstracting an idea through the \textit{Similar} button (Figure~\ref{fig:similar}a\&b). When clicked, \sys prompts the LLM to identify the key categories that the selected idea represents (see Prompts in Appendix~\ref{prompt_abstract_generation}--\ref{prompt_abstract_check}). The returned categories are compared with the existing set, merged when appropriate, and added when new. Selecting one of these categories creates a schema card and three associated solutions, which forms a new branch in the multi-thread view. For instance, in the example scenario, the user who began with lemon spray as a chemical treatment could pivot to its targeted area of use, leading to ideas such as a pen-style applicator or a laser-focused foam(see figure~\ref{fig:similar}). Choosing one of these adds it to the shared category set, which remains available in the sidebar for any canvas and continues to support the generation of new branches.

Because both the initial seeds and the Similar feature rely on the same category structure for generating new idea branches, users can move smoothly across threads and expand the space along the chains. They can switch between ideas, compare parallel paths, and generate new directions while keeping existing \tradeoffmitigation chains intact and easy to revisit. This design supports broad exploration while preserving the clear traces needed for deep refinement.

\subsection{Designing for Human Agency}
Our third design goal is to support a human–AI partnership model where AI probes and extends users’ reasoning without deciding for them. To achieve this, \sys uses a tree structure on the canvas that is shaped jointly by human actions and AI suggestions but remains under user control. When users generate a tradeoff or solution, the system links the new node to the currently selected one to record the user’s line of thought, but users can freely move, insert, or delete nodes at any time and in any place. This flexibility allows each branch to grow according to the user’s decisions rather than a fixed system logic. The Auto Layout button (see Figure~\ref{fig:interface_overview}) offers optional organization support, and all seed ideas remain accessible in the sidebar for switching between trees.

To help users bring their own reasoning into the process, \sys provides lightweight ways to add content and ask questions directly on the canvas. Users can add tradeoffs with the \textit{+ Tradeoff} button and add solutions with the \textit{+ Solution} button on any card. They can also pose free text questions about any node (Figure~\ref{fig:similar}f). The LLM answers in the context of the selected idea, the design brief, and the user’s question (see Prompt in Appendix~\ref{prompt_answer_gen}). All responses appear in place so users receive support without leaving their current thread. Together, these choices keep users in control of how branches develop while allowing AI to assist in a way that sustains engagement.

\section{Evaluation}
We evaluated whether \sys’s multi-threaded \tradeoffmitigation design improves both ideation outcomes and process. For outcomes, we examined whether \sys helps users generate higher-quality ideas. For process, we focused on how widely participants explored the design space (breadth), and how far they developed individual idea branches (depth). These dimensions are central in early exploration~\cite{tversky2010creativity} and they directly reflect our design goals. 
% The \tom chain was created to support depth by helping users articulate trade-offs and push ideas forward, while the multi-threaded structure was created to support breadth by enabling branching, comparison, and navigation across alternatives. 
To investigate these questions, we conducted a mixed-methods, within-subjects study with 20 design and engineering professionals who solved two open-ended problems using both FlexMind and a ChatGPT baseline. A follow-up case study with 6 senior practitioners examined FlexMind’s real-world applicability (see Section~\ref{expert_section}).

\subsection{Controlled User Study}
In this controlled study, we tested whether FlexMind’s multi-threaded \tradeoffmitigation structure improves ideation outcomes and exploration behaviors beyond commonly used ideation tools. We selected ChatGPT as the baseline as it is the widely used AI tool for everyday ideation, which maximizes ecological validity~\cite{sinlapanuntakul2025impacts}. A familiar tool also minimizes cognitive burden for participants, preventing learning difficulty from influencing baseline performance. Although other baselines such as different LLMs or search-augmented tools are possible, they would reduce comparability and introduce unnecessary confounds. Our goal was to compare FlexMind’s scaffolds against the most common status quo. To preserve natural ideation behavior in the baseline, we kept prompting fully open-ended rather than constraining participants with predefined templates or workflows. This allowed participants to use ChatGPT as they normally would, avoiding artificial limits on performance. To ensure fairness between conditions, we held the task and time constant and used the same underlying model (\texttt{GPT-o4-mini}) in both conditions.

\subsubsection{\textbf{Design Problems}}

Participants addressed two open-ended design problems: (1) how to clean laundry using less water and (2) how to minimize accidents caused by walking while texting (see full prompts in Appendix~\ref{apd_task_description}.). Problem 1 is from a real-world creative problem-solving challenge\footnote{https://www.mindsumo.com/contests/dry-laundry}, and problem 2 is from prior work~\cite{ma2023conceptual} used for LLM creativity-related evaluation. We selected these problems because they invite diverse solution strategies, and require evaluating trade-offs and refinements. Both tasks were pilot-tested to calibrate difficulty and timing. 

\subsubsection{\textbf{Participants}}

We recruited 20 design and engineering professionals (10 female, 10 male), each with at least one year of professional experience and a degree in design or engineering. None had previously worked on the two problems used in the study. The sample balanced backgrounds (10 engineering, 10 design). We sampled working design/engineering professionals to ensure professional fluency and reduce domain-specific bias in ideation practices. Recruitment was conducted via posts to message-board channels (e.g., Slack) at several U.S. universities. Participants provided  consent under IRB approval and received \$35 compensation. 

\subsubsection{\textbf{Procedure}}

Participants’ goal in each task was to generate ideas for the assigned design problem. To observe their exploration and refinement strategies, we asked them to ideate as they normally would while thinking aloud. Each participant completed both tasks, one with FlexMind and one with the ChatGPT baseline, with task–condition order counterbalanced. Sessions lasted about 90 minutes (remote via Zoom) and followed a fixed sequence: orientation and instructions, a 30-minute ideation period for the first task, an immediate post-condition survey, a 30-minute ideation period for the second task under the alternate condition, and a short semi-structured interview. Before starting the FlexMind condition, participants watched a tutorial and completed a brief practice task (“design a lightweight exercise device that collects energy from human motion”) to learn the system’s features. In both conditions, participants could consult the web for quick fact checks but were asked to rely only on the assigned tool (FlexMind or ChatGPT) to generate, organize, and develop ideas. To standardize deliverables for blinded expert evaluation, participants recorded the ideas they judged worth keeping in a structured spreadsheet after each condition.

The post-condition questionnaire captured participants’ experience with that tool. It assessed the ideation process and perceived support, self-evaluation of the final deliverable, and the Creativity Support Index~\cite{cherry2014quantifying}. In the FlexMind condition, participants also rated the usefulness of core features such as externalized thinking flows, the initial overview page, tradeoff generation, solution generation, adding tradeoffs and solutions, and Q\&A. This procedure preserved ecological validity by allowing naturalistic ideation with familiar tools while still producing comparable outputs and analyzable process traces through think-aloud and interaction logs.

\subsection{Measurement}

We evaluate FlexMind on what people produce and how they explore. To ensure the validity of idea-quality assessment, we recruited senior practitioners to rate anonymized outputs, whose evaluations provide domain-informed judgments of ideas. To assess exploration behavior, we analyzed the participants’ interactions with both systems. FlexMind’s logs directly encode branches and \tradeoffmitigation chains, while baseline chats unfold linearly. We therefore reconstructed comparable exploration trees from ChatGPT transcripts through manual annotation, which allowed us to measure breadth, depth, and evaluation moves in a consistent way across conditions. We also collected post-condition surveys to capture perceived support, complementing the behavioral measures and expert evaluations.

\subsubsection{Idea Quality Evaluation.} 
Our goal is to test whether FlexMind produces higher-quality ideas than the baseline. Following creativity studies~\cite{shah2003metrics,ma2023conceptual}, we decompose quality into novelty, feasibility, and value to align the evaluation with real-world R\&D decision criteria. We define \textbf{novelty} as the degree to which an idea is not only rare but also ingenious or surprising; \textbf{feasibility} as implementability with current resources and technology; and \textbf{value} as the extent to which the idea addresses the problem effectively. Consistent with prior work, we aggregate these dimensions using the geometric mean to produce a single quality score~\cite{shah2003metrics}.

To obtain judgments that reflect professional standards rather than self- or crowd-perception, we recruited four domain experts to rate the submitted ideas, blind to condition. Two experts rated Task 1 (reducing water use in laundry) and two rated Task 2 (minimizing accidents from walking-while-texting) . Task 1 raters were engineering and R\&D specialists with hands-on product implementation experience: one is the head of R\&D at a major consumer appliance manufacturer, and the other is a faculty member in the mechanical engineering department at a U.S. university. Task 2 raters were engineering design specialists with practical mobility-project experience: one is a faculty member at a U.S. university, and the other is a senior industry researcher. Both have research focuses on accessibility. We compensated each expert with \$200 for their effort. The experts met with the first author to calibrate the rubric using 22 randomly selected ideas from each task. After consensus on operational definitions and examples for novelty, feasibility, and value, the experts independently rated the remaining anonymized ideas. Inter-rater reliability, computed as ICC(2,k), indicated good agreement for \textbf{novelty} and \textbf{feasibility} and moderate--good for \textbf{value} (Task1/Task2: \textbf{novelty}~0.77/0.77; \textbf{feasibility}~0.76/0.79; \textbf{value}~0.56/0.70; full table in Appendix~\ref{ICC}). For each idea, we averaged the two raters’ scores on each dimension and then computed the geometric mean to yield the overall idea-quality score used in analysis.

\subsubsection{Exploration Behavior}
\label{tree_method}
\begin{figure}[t]
  \centering
  \includegraphics[width=\linewidth]{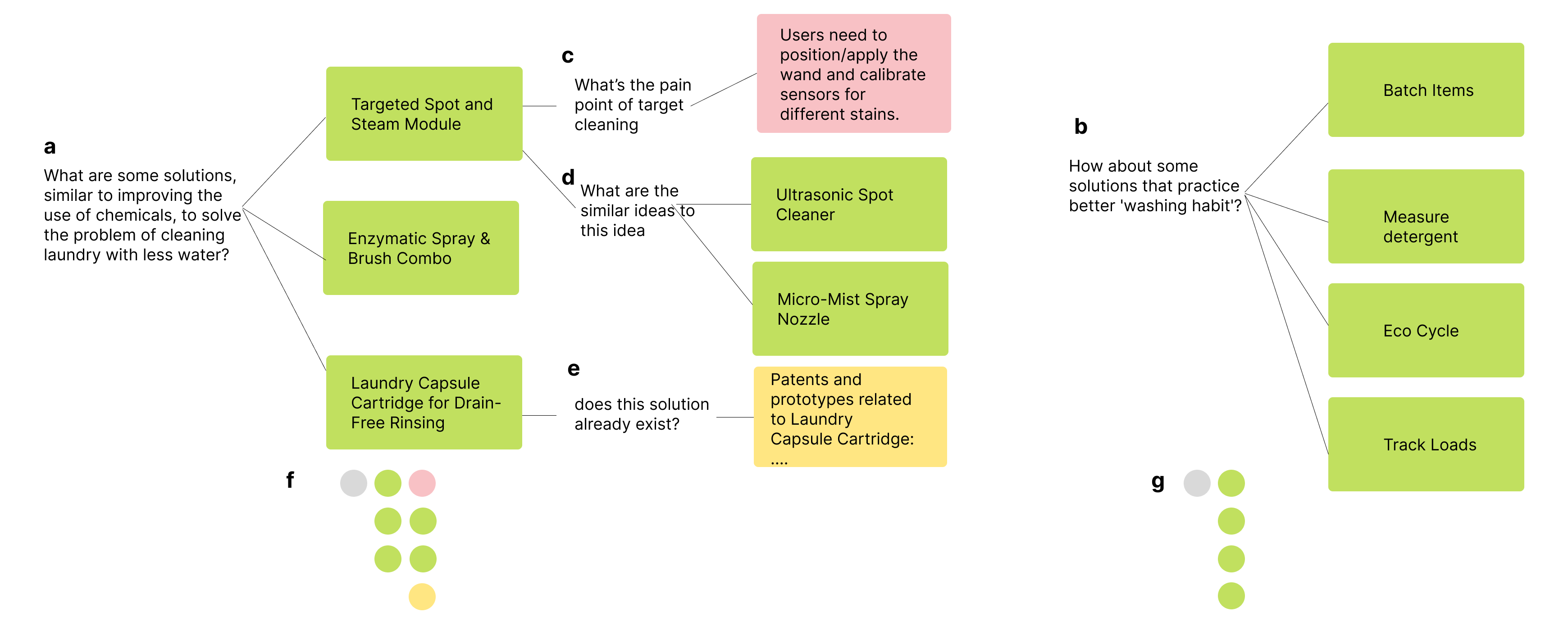}
  \caption{\textbf{Annotation example in the baseline condition.} In the baseline (ChatGPT) condition, user actions and generated information are  reconstructed into trees. Green cards represent solutions, red cards represent trade-offs, and yellow cards represent other knowledge. Nodes \textbf{(a)} and \textbf{(b)} show two trees initiated from user prompts for solutions. Node \textbf{(c)} shows a user asking ChatGPT to analyze trade-offs, while node \textbf{(d)} shows a request for similar solutions. Node \textbf{(e)} illustrates a user asking follow-up questions on prior information. Node \textbf{(f)} and \textbf{(g)} show abstract visualizations of the same trees, indicating how nodes and branches are counted.}
  \Description{Two reconstructed idea trees are shown. On the left, panel \textbf{(a)} begins with a user prompt asking for solutions, which generates three green solution cards: “Targeted Spot and Steam Module,” “Enzymatic Spray \& Brush Combo,” and “Laundry Capsule Cartridge for Drain-Free Rinsing.” From these nodes, one red card (c) represents a trade-off (“Users need to position/apply the wand and calibrate sensors for stains”), while green and yellow cards (d, e) represent similar solutions and additional knowledge. On the right, panel \textbf{(b)} shows another user prompt generating green solution cards such as “Batch Items” and “Eco Cycle.” At the bottom of each tree (f, g), abstract circular diagrams summarize node and branch counts. Green denotes solutions, red denotes trade-offs, and yellow denotes other knowledge. The figure illustrates how baseline conversations with ChatGPT were annotated into structured trees for comparison with \sys.}
  \label{fig:annotation example}
\end{figure}

To evaluate whether \sys fosters broader and deeper exploration than the baseline, we first needed a common analytic representation. \sys natively records a branching tree of explored information, while the baseline produces a linear chat transcript. Because these formats are not directly comparable, we constructed each baseline session into a tree so that breadth and depth could be measured consistently across both conditions. We synchronized three data sources for every session: think-aloud protocols, interaction logs, and screen and audio recordings. Following prior work that models thinking as linked action chains~\cite{gero2020framework}, we coded the sequence of user actions and the information generated from those actions. We extracted “thinking threads” and represented them as branches, with different ideas forming branches and distinct themes forming separate trees (see Figure~\ref{fig:annotation example}).

In the baseline condition, each user move was labeled with one of the following categories (definitions in Appendix~\ref{apd_codebook}): \emph{analyzing idea trade-offs} (independently or with ChatGPT/search), \emph{generating solutions to the trade-offs} (with ChatGPT/search), \emph{finding solutions to trade-offs} (independently or with ChatGPT/search), \emph{finding similar ideas} (independently or with ChatGPT/search), \emph{asking for background knowledge}, \emph{adding self-generated ideas}, and \emph{other prompts} such as directly asking for solutions to the problem. Four authors refined the criteria together. Two authors annotated all sessions jointly, resolving disagreements through discussion. We represented each session using two  node types:
(1) \emph{Action nodes} capture the participant (or tool) actions, such as analyzing trade-offs or generating mitigations
% , find similar ideas, request background knowledge, or add a self-generated idea; 
and (2) \emph{Information nodes} capture  action outputs, such as  ideas or trade-offs.
%, mitigations, or background facts.
% This separation (i) enables reconstruction from linear baseline transcripts in a way comparable to \sys’s structured logs, (ii) aligns what is rendered across conditions (\sys visually organizes information as a tree while recording actions in logs), and (iii) yields stable analytic units—actions and the information they produce—so downstream measures apply consistently. 
This enables reconstructing baseline logs in a form comparable to \sys’s logs for downstream measures.
For example, if a participant prompted ChatGPT for solutions and it returned four ideas, four information nodes were linked to an action node representing that prompt (Figure~\ref{fig:annotation example}b). We linked actions to prior information nodes when (1) two actions addressed the same design idea or (2) a later action developed the output of an earlier one, following the   logic~\cite{gero2020framework}.

Although \sys visually renders only information nodes, it logs all user actions. For consistency across conditions, we counted only information nodes in both conditions. In \sys, we further excluded question nodes (e.g., Figure~\ref{fig:similar}e) to match what is counted in the baseline. This yields comparable measures of the number of distinct information elements explored and the amount of prior context available to each node. Full annotated trees are shown in Figure~\ref{fig:all_tree}. While a web browser was available in the \sys condition, only 2 of 20 participants used it, each fewer than three times for brief fact checks. To focus analyses on in-system behavior, browser search logs were excluded from \sys, whereas ideas generated via search were retained for analysis in the baseline, where search was used more frequently.

\subsubsection{\textbf{Tree measurement}}
We used the reconstructed trees to quantify \emph{breadth} and \emph{depth} of exploration. Breadth was measured by (i) the number of trees, which reflects distinct themes, and (ii) the number of information nodes, which reflects the overall scope of explored content. Depth was measured by (iii) branch length and (iv) the longest path within each tree, which indicate how far a line of thought was pursued. For submitted ideas, we computed idea chain length as the path from the root to the information node that produced the idea. These metrics allow us to test whether \sys produces more trees, more explored content, and deeper branches than the baseline, and to examine differences by action category. To measure how people navigated among their thinking threads and whether they actively reused prior information or simply discarded it and generated new ideas, we defined a `jump' measure. We categorized jumps into five types:
\begin{itemize}
    \item \textbf{New Tree}: starting a new tree (e.g., from action~a to action~b in Figure~\ref{fig:annotation example}).  
    \item \textbf{Switch tree}: switching to a different existing tree (e.g., from action~b to action~c in Figure~\ref{fig:annotation example}).  
    \item \textbf{Continue branch}: continuing actions along the same branch of a tree (e.g., if~c directly follows action~a, then from action~a to action~c in Figure~\ref{fig:annotation example}).  
    \item \textbf{Parallel branch}: actions that share the same parent node within a tree.  
    \item \textbf{Cross branch}: actions that are neither siblings nor in the same branch as the previous action but still occur in the same tree (e.g., from action~d to action~e in Figure~\ref{fig:annotation example}).  
\end{itemize}

The distribution of jump types reveal participants’ behavioral patterns. A higher percentage of \textit{first action} jumps suggests shallow engagement, where participants may discard or adopt ideas too early. In contrast, a higher percentage of \textit{same branch} jumps indicates sustained reasoning within a line of thought. We also examined whether participants engaged with \sys outputs rather than skimming. System responses averaged 9.33,s (SD = 2.22; $n=20$ actions), while consecutive user actions averaged 84.54,s apart (SD = 60.18), and 88.27\% of intervals exceeded three times the system’s response time. This pattern suggests users paused to consider content rather than clicking through superficially.

These choices place both conditions on equal footing. Both are represented as trees, the same information-node units support all metrics, and browser activity is controlled.
%The Results section therefore tests, on comparable structures and metrics, whether \sys expands the \emph{breadth} (more trees, more information nodes) and \emph{depth} (longer branches) of exploration, and which action categories most contribute to any observed differences.

\subsubsection{\textbf{Analysis Methods}}
\label{controlled_study_method}
We use the following methods for analyzing our user study results. 
\begin{enumerate}
    \item Idea quality (quant). Expert ratings on novelty, feasibility, and value were combined via the geometric mean. We compared average ratings across conditions and fit a linear mixed-effects model predicting average score from condition, with participant as a random intercept. Rubric-discussion items ($n=22$ total; 11 from each condition) and ideas judged too vague by experts (baseline: n=9, \sys: n=2) were excluded before modeling.
    \item Exploration behavior (quant). We built comparable information-node trees using the action/information codebook(see ~\ref{tree_method}), then tested breadth, depth, and jump patterns (Wilcoxon signed-rank with Bonferroni). A linear mixed-effects model predicted overall idea quality, with condition (baseline vs. \sys) and chain length as fixed effects and participant as a random intercept. Because some participants made edits to their submitted ideas rather than using the exact wording provided by \sys or ChatGPT, tracing every idea back to its corresponding information node would have required intensive manual work. To balance accuracy with feasibility, we randomly sampled two submitted ideas per participant per condition. Using the established baseline annotations, two authors coded the idea chain length. This sampling did not cover all ideas but ensured inclusion of both directly represented and indirectly linked ideas while keeping the analysis tractable and reliable. 
    \item Surveys (Likert) \& CSI. Post-condition questionnaires included Likert-type items on experience/support and the Creativity Support Index~\cite{cherry2014quantifying}; because the study was single-user, the CSI Collaboration factor was removed. We compared ratings with Welch’s $t$ (Bonferroni corrections are applied where applicable).
    \item Qualitative analysis (think-aloud + interview). We analyzed think-aloud protocols and post-task interviews using inductive thematic analysis. Themes were triangulated with interaction logs and tree structures to connect verbalized reasoning with observed exploration behavior.
%(constant-comparison across participants; memoing to surface mechanisms).
\end{enumerate}

\section{Results}
We found that participants in the \sys condition generated higher-quality ideas and explored each idea more deeply, without sacrificing overall exploration breadth. Deeper exploration was also associated with higher-quality ideas. This balance of depth and breadth emerged from the multi-threaded \tradeoffmitigation structure in \sys.

\subsection{Better Ideas and More Active Thinking}
\subsubsection{Idea Quality Improvement}
\begin{figure}[t]
  \centering
  \includegraphics[width=0.6\linewidth]{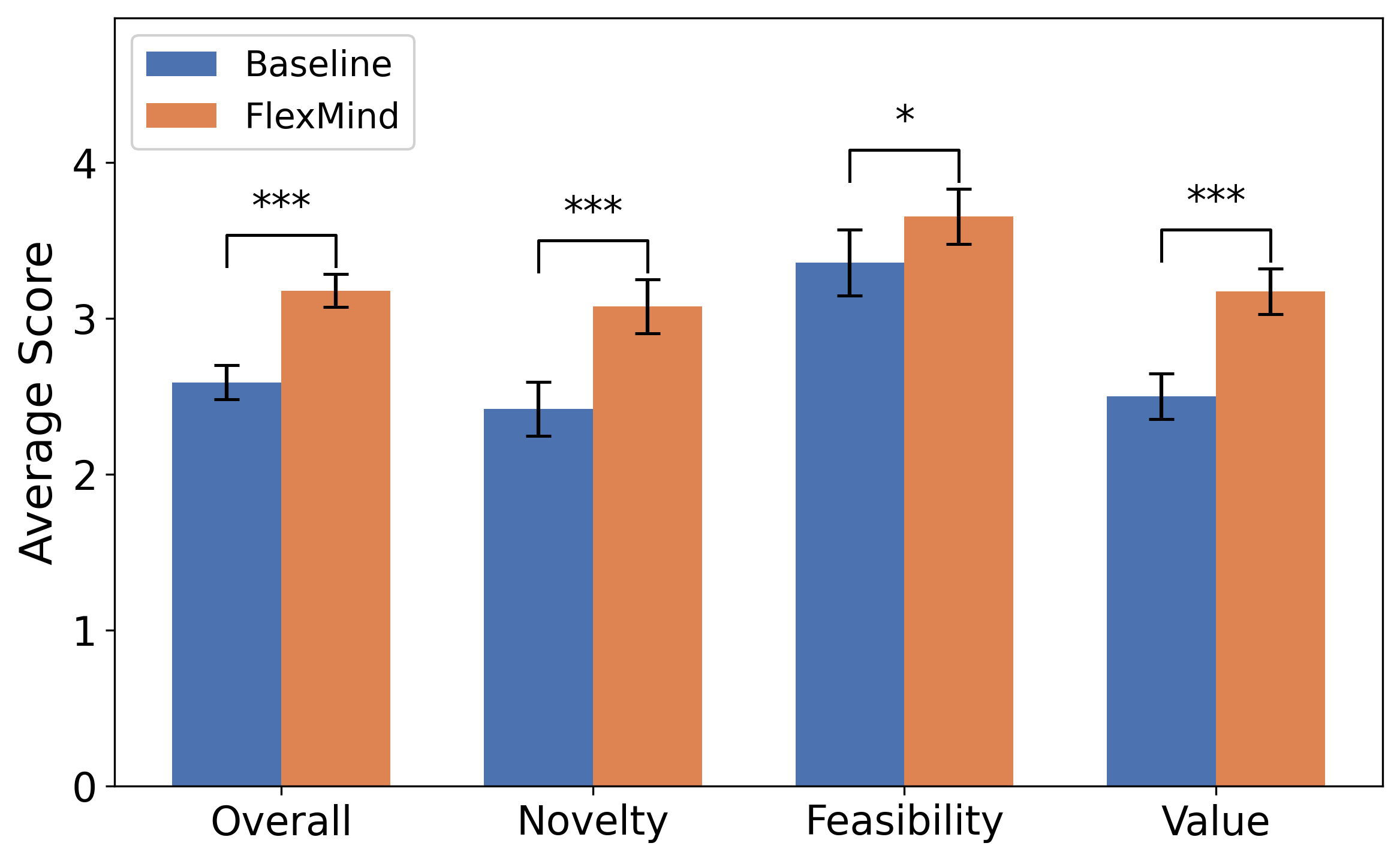}
  \caption{\textbf{Experts' average idea ratings for ideas in the baseline ($n=140$) and FlexMind ($n=130$).}  FlexMind ideas received significantly higher scores overall, with improvements in novelty($p<.001$), value ($p<.001$) and feasibility ($p<.05$).}
  \Description{A bar chart compares expert ratings of ideas across two conditions: baseline (blue bars) and FlexMind (orange bars). Four groups are shown on the x-axis: Overall, Novelty, Feasibility, and Value. For all four metrics, FlexMind bars are higher than baseline. Statistical significance markers are shown above: three asterisks (**) above Novelty and Value indicate $p<.001$, and one asterisk () above Feasibility indicates $p<.05$. The y-axis shows average score from 0.0 to 4.0. Error bars represent standard errors.}
  \label{fig:expert_ratings}
\end{figure}

Overall, participants in the \sys condition submitted significantly higher-quality ideas (see Figure~\ref{fig:expert_ratings}) than the baseline. We found that, within participants, ideas quality was significantly higher when using \sys ($M = 3.18, SD = 0.63$) than in the baseline condition ($M = 2.59, SD = 0.65$; Welch’s t-test, $p < 0.001$). This finding was further substantiated with a linear mixed-effects model, which controlled for participant-level variance, similarly indicating a significant positive effect of the \sys condition on idea quality scores (β = 0.47, p < .001).

\begin{figure}[t]
  \centering
  \includegraphics[width=0.6\linewidth]{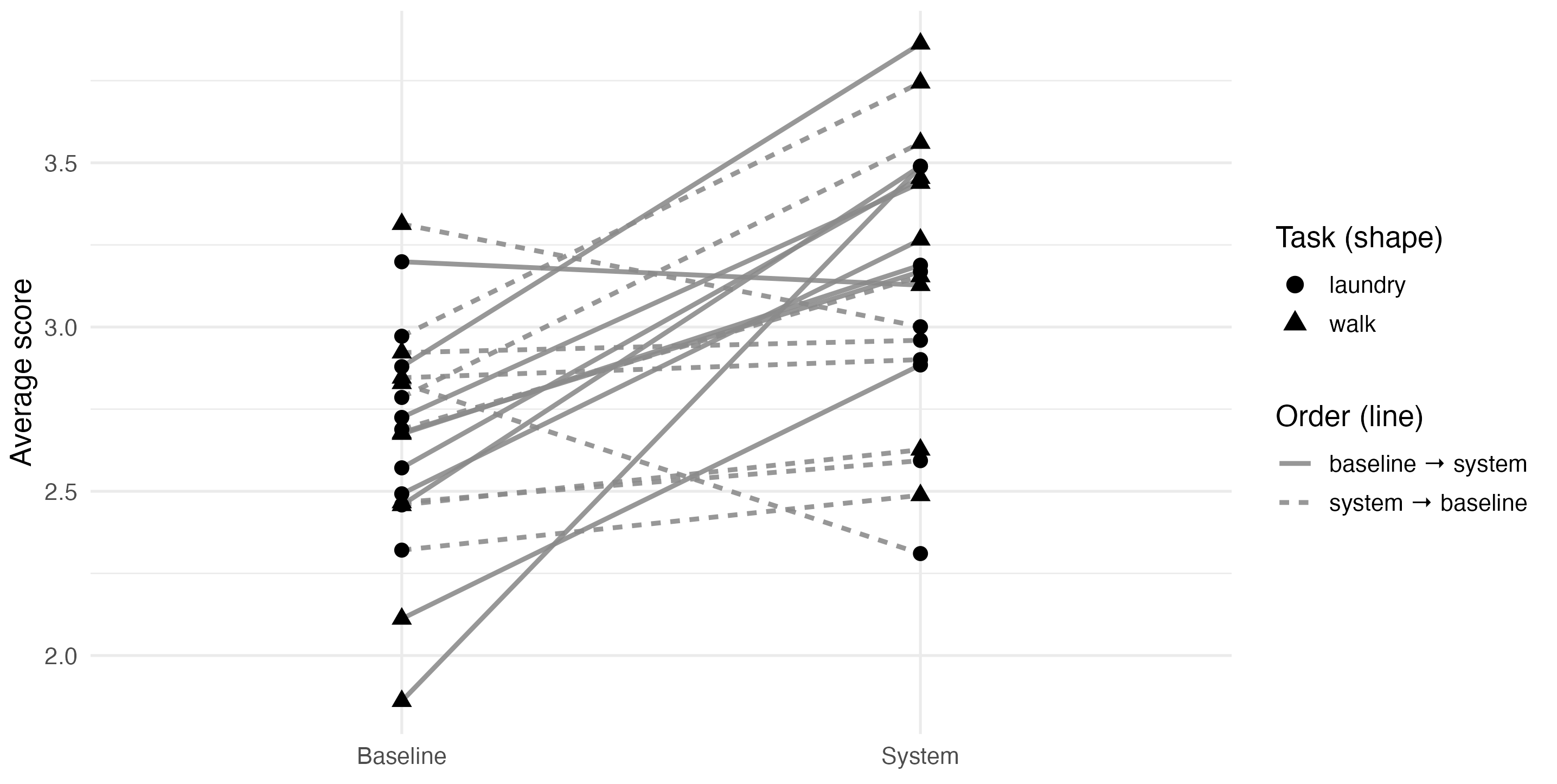}
  \caption{\textbf{Participant-level expert ratings across conditions.} Each point is a participant’s mean expert rating of all their submitted ideas (1–5) in the Baseline and \sys conditions; lines connect the same participant. Line type encodes study order (solid = baseline $\rightarrow$ system; dashed = system $\rightarrow$ baseline). Point shape encodes task ( $\bullet$ laundry; $\blacktriangle$ walk). Most lines rise from Baseline to System, indicating higher scores with \sys for 17 of 20 participants (3 decreased).}
  \Description{Slope chart with two columns labeled Baseline (left) and System (right). Each participant appears as two points (one per column) connected by a line; the y-axis is mean expert rating from 1 to 5. Line type encodes study order: solid = baseline → system; dashed = system → baseline. Point shape encodes task: ● laundry, ▲ walk. Most lines rise from Baseline to System, indicating higher scores with \sys; specifically 17 of 20 increase, 3 decrease}
  \label{fig:all_score}
\end{figure}

The positive effect of \sys on idea quality was broadly consistent across participants; as illustrated in Figure~\ref{fig:all_score}, 17 of the 20 participants had higher average scores in the \sys condition. Examining the distribution of scores, a large majority of the top-third ideas (76.4\%) originated from the \sys condition, while the majority of the bottom-third ideas (71.9\%) came from the baseline. As shown in Figure~\ref{fig:expert_ratings}, this pattern of improvement was also reflected in the individual metrics of novelty, feasibility, and value, all of which were significantly higher in the \sys condition. Table~\ref{table:score-examples} shows  illustrative examples of ideas across different score ranges.

% Participants’ reflections aligned with these quantitative gains. Several noted that \sys’s outputs were higher quality and more actionable; for instance, P13 said, \pquote{I feel this system was much better at giving the solutions. The solutions it provided were more in depth… it gave me a good idea as a starting point for the process.} Others emphasized that \sys helped them improve their own ideas, with P16 noting, \pquote{During the ideation process the system helped me structure my thinking… I was able to see when the ideas were not good, and move quickly through bad ideas and good ideas.}

\begin{table}[!ht]
\centering
\small
\renewcommand{\arraystretch}{1.1}
\setlength{\tabcolsep}{6pt}
\begin{tabular}{p{0.18\linewidth} p{0.27\linewidth} p{0.27\linewidth} p{0.18\linewidth}}
\hline
\textbf{Score Range} & \textbf{Baseline Example} & \textbf{\sys Example} & \textbf{Idea count (blue: baseline, orange: \sys)} \\
\hline
Low (1.000--2.621) & 
Devices do not use normal display -- flashing screen (Task 2) & 
Hang shirts and jeans out in fresh air after a single wear—often airing out is enough to eliminate odors without any water. (Task 1) &
\useimage{ }{ }{low.png} \\
Medium (2.639--3.271) & 
A new low-weight cycle on the machine that you can input how many clothes, and it uses less water (Task 1) & 
Wearable vest with multiple vibration points that provide directional feedback about nearby hazards. Different vibration patterns indicate various types of obstacles or dangers. (Task 2)&
\useimage{ }{ }{medium.png} \\
High (3.302--4.481) & 
Foam-Flush Washer Add-On: A clip-on canister for front-loaders that injects dense detergent foam into the drum. The machine tumbles as usual but only mists in enough water to rinse out bubbles. (Task 1) & 
Gesture-Required Activation, which requires specific hand gestures or phone movements to confirm the user is stationary before allowing text input. Prevents accidental activation while walking. (Task 2) &
\useimage{ }{ }{high.png} \\
\hline
\end{tabular}
\caption{\textbf{Examples of ideas at different quality score ranges.} For each score band (Low, Medium, High), we show representative ideas from Baseline and \sys, note the task (Task 1: laundry; Task 2: walking/texting), and report how many ideas fall in each band (right-hand bars; blue = Baseline, orange = \sys). Participants submitted more high-quality ideas with \sys (High: 68 vs. 21) and fewer low-quality ideas (Low: 27 vs. 69); Medium counts were similar (42 vs. 46).} 
\label{table:score-examples} 
\end{table}

\begin{table}[ht]
\centering
\small
\renewcommand{\arraystretch}{1.2}
\setlength{\tabcolsep}{6pt}
\begin{tabular}{lrrrrrr}
\hline
\textbf{Term} & \textbf{Coef.} & \textbf{Std.Err.} & \textbf{z} & \textbf{p\,$>$\,|z|} & \textbf{[0.025} & \textbf{0.975]} \\
\hline
Intercept & 2.663 & 0.086 & 31.05 & < 0.0001 & 2.494 & 2.831 \\
C(condition)[T.system] & 0.473 & 0.115 & 4.10 & <0.0001 & 0.247 & 0.699 \\
Group Var & 0.104 & 0.277 & 0.38 & 0.707 & -0.439 & 0.648 \\
\hline
\end{tabular}
% \caption{Mixed-effects regression results. Formula: average\_score $\sim$ C(condition)}. 
\caption{\textbf{Mixed-effects regression results predicting idea quality score (1–5). }The model ($average\_score \sim condition$) shows that ideas produced using \sys received significantly higher ratings than the baseline by +0.47 points on a 1–5 scale (SE = 0.12, $z$ = 4.10, $p < .0001$, 95\% CI [0.25, 0.70]). The intercept (2.66) represents the estimated mean score for the baseline condition.}
% In-sample RMSE = 0.3400, Pseudo $R^2$ = 0.4090.
\label{table:mixed_effects_overall}
\end{table}

\subsubsection{Increased Active Engagement and Evaluation Behavior in the Ideation Process}
\begin{figure}[t]
  \centering
  \includegraphics[width=0.8\linewidth]{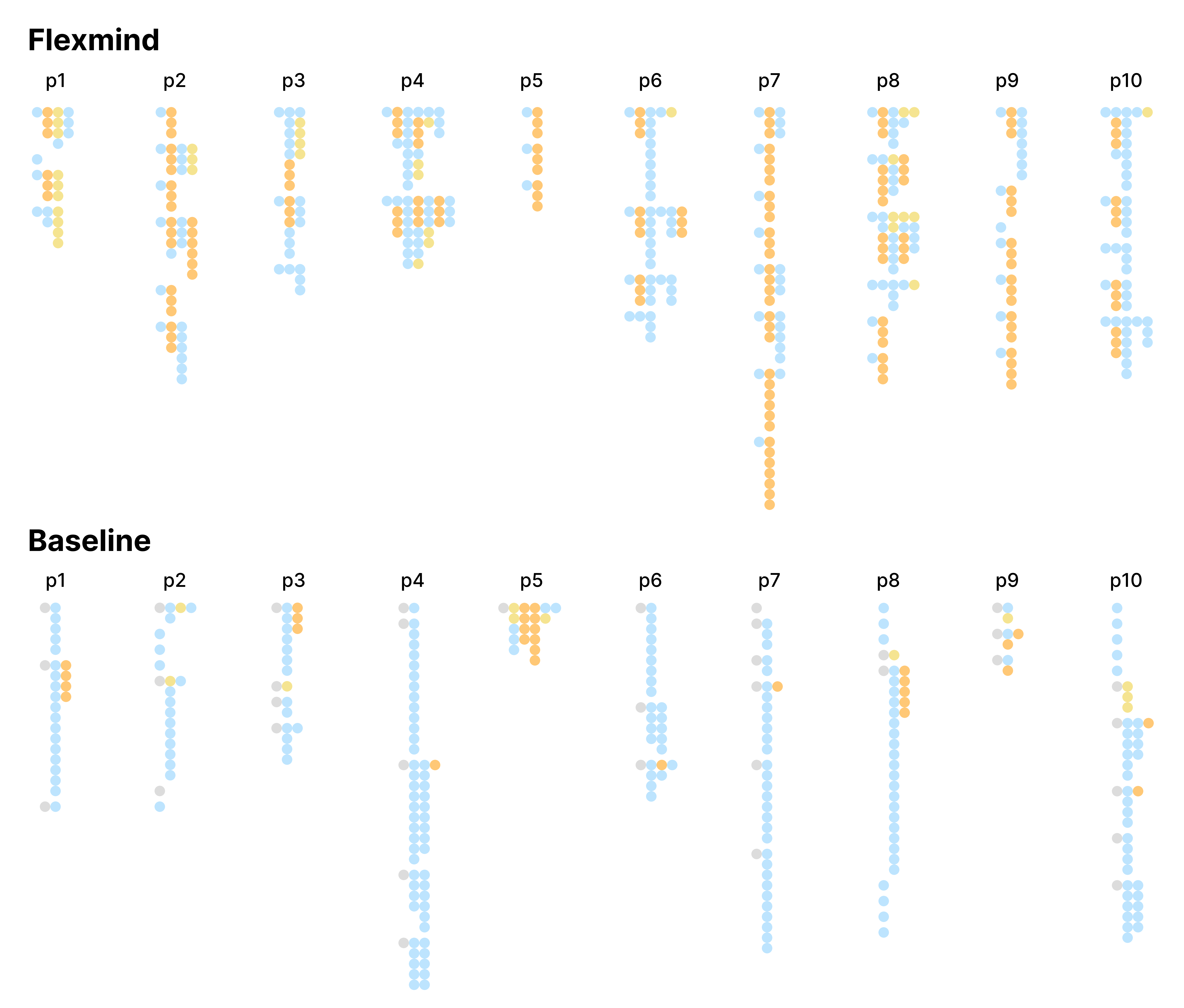}
    \caption{\textbf{Participants’ Ideation Processes by Condition.} For participants P1--P10, FlexMind (top) and Baseline (bottom) sessions are rendered as trees (See the data for other participants in the Appendix~\ref{all_tree_2}). The \textbf{x}-axis encodes \emph{depth} (how deep a participant pursued an idea within a tree using the tradeoff-mitigation chains), and the \textbf{y}-axis encodes \emph{breadth}, both within trees (number of branches) and across trees (number of distinct trees). Each dot is a node: blue = solutions, orange = trade-offs, yellow = other information (i.e., answers to user questions); in Baseline, gray dots depict the initial prompt/search query used to start a tree (not counted as tree nodes). Overall, more participants in FlexMind explored deeper trees and documented more trade-offs than in the Baseline.}

    \Description{Two rows of small node-and-branch “trees,” one labeled “FlexMind” (top) and one labeled “Baseline” (bottom). Each row shows ten columns labeled p1–p10. Within each participant column are one or more compact trees made of colored dots connected in simple branches. Blue dots mark solution nodes, orange dots mark trade-off nodes, and yellow dots mark other information (such as questions or similar-by-schema ideas). In the Baseline row, light gray dots appear at the tops of some trees to indicate the initial prompt or search query; these gray dots are not counted as tree nodes. Compared to Baseline, the FlexMind row shows more and taller trees with more branching and more orange nodes.}
      
  \label{fig:all_tree}
\end{figure}

Participants performed significantly more actions in the \sys condition (M = 20.40, SD = 6.64) than in the baseline (M = 9.28, SD = 3.27; $t(19) = 7.75, p < .001$). More importantly, a larger share of these actions were trade-offs for idea evaluations (see Figure~\ref{fig:all_tree} and Table~\ref{table: action_type}). Baseline sessions were dominated by direct solution-seeking, with 81\% of “other” actions consisting of simple prompts for more ideas or web searches. The pattern shows that multi-threaded \tradeoffmitigation structure encourages users to examine limitations and iterate, rather than repeatedly asking for new ideas. The higher frequency of trade-off analysis therefore indicates that FlexMind successfully prompted deeper engagement with emerging ideas.

These results suggest that \sys helped participants thinking more actively and generate higher quality ideas. In the sections below we further investigate these results to understand how participants' usage of \sys influenced the depth and breadth of exploration as well as its relationship with idea quality.

\begin{table}[!ht]
    \centering
    \begin{tabular}{l l l l}
    \hline
        Action & \sys (M$\downarrow$, SD)  & Baseline Mean(SD) & W (Bonf. p) \\ \hline
        search for or generate tradeoff & 5.60 (2.39) & 0.50 (1.05) & 0.0 (0.001)\textsuperscript{*} \\
        add solution of their own & 4.55 (2.89) & 3.35 (3.53) & 71.5 (1.000) \\
        search for or generate solution (for sub problems or tradeoffs) & 3.50 (2.59) & 0.50 (0.89) & 6.0 (0.003)\textsuperscript{*} \\
        ask fact-check question & 3.00 (2.79) & 1.50 (1.69) & 26.5 (0.335) \\
        search for or generate similar ideas based on seed ideas & 2.25 (2.47) & 0.75 (1.21) & 3.5 (0.017)\textsuperscript{*} \\
        add tradeoff of their own & 1.50 (2.19) & 1.65 (1.84) & 57.5 (1.000) \\
        other & N/A & 1.03 (1.46) & N/A \\ \hline
\end{tabular}
    \caption{\textbf{Participants’ actions with FlexMind vs.\ baseline.} FlexMind led to significantly more searching or generating (i) \emph{trade-offs}, (ii) \emph{solutions for sub-problems/trade-offs}, and (iii) \emph{similar ideas based on seed ideas} (Wilcoxon signed-rank with Bonferroni correction: \(p = .001, .003, .017\), respectively). Other action frequencies did not differ significantly. Values are means per participant with SD in parentheses. \(W\) is the Wilcoxon statistic; Bonferroni-corrected \(p\)-values appear in parentheses; * indicates significance after Bonferroni correction. \textit{N/A} indicates the category was not used/available in that condition.}

    \label{table: action_type}
\end{table}

\subsection{Deeper Exploration}
The idea trees within the \sys condition were significantly deeper than in the baseline condition(see Table~\ref{table:tree}). In the baseline, nearly half of all trees were single-node (49\%), indicating ideas that were recorded but not developed. In contrast, only 13\% of trees in \sys were single-node, showing that participants extended their ideas far more often when supported by the system. Longest-path depth was also significantly greater in \sys (M = 2.37) than in the baseline (M = 1.59; Wilcoxon W = 27.0, Bonf. p < .05). Importantly, this increase in depth did not reduce breadth; the average number of trees explored was similar across conditions (Baseline: 5.95, \sys: 5.55, Table~\ref{table:tree}). Together, these results show that \sys encouraged users to elaborate ideas rather than simply move on, achieving a core design goal of deeper reflective reasoning within each branch.

%Participants in \sys explored a similar number of trees compared to the baseline (see Table~\ref{table:tree}).
%However, the threads in the \sys condition were more elaborate. In both conditions, some trees contained only a single node. A one-node tree occurs when a participant records an idea without developing it further, typically reflecting shallow engagement. In contrast, multi-node trees capture idea elaboration and extension via tradeoffs, solutions, and questions. In the baseline condition, 49\% of the trees were single-node trees, whereas in the \sys condition, only 13\% were single-node trees. The average branch length in \sys was longer than in the baseline, though the difference was not significant. However, tree depth, measured as the average length of the longest chain in the trees, was significantly higher in the \sys condition (M = 2.55) compared to the baseline condition (M=0.86). Taken together, these results suggest that participants not only engaged with more thinking threads in the \sys condition, but also carried out more steps along those threads. This indicates \textit{deeper} engagement with the provided ideas, rather than shallow exploration.

\subsubsection{Longer chains are associated with better idea quality}
\label{quality_analysis}
\begin{figure}[t]
  \centering
  \includegraphics[width=0.6\linewidth]{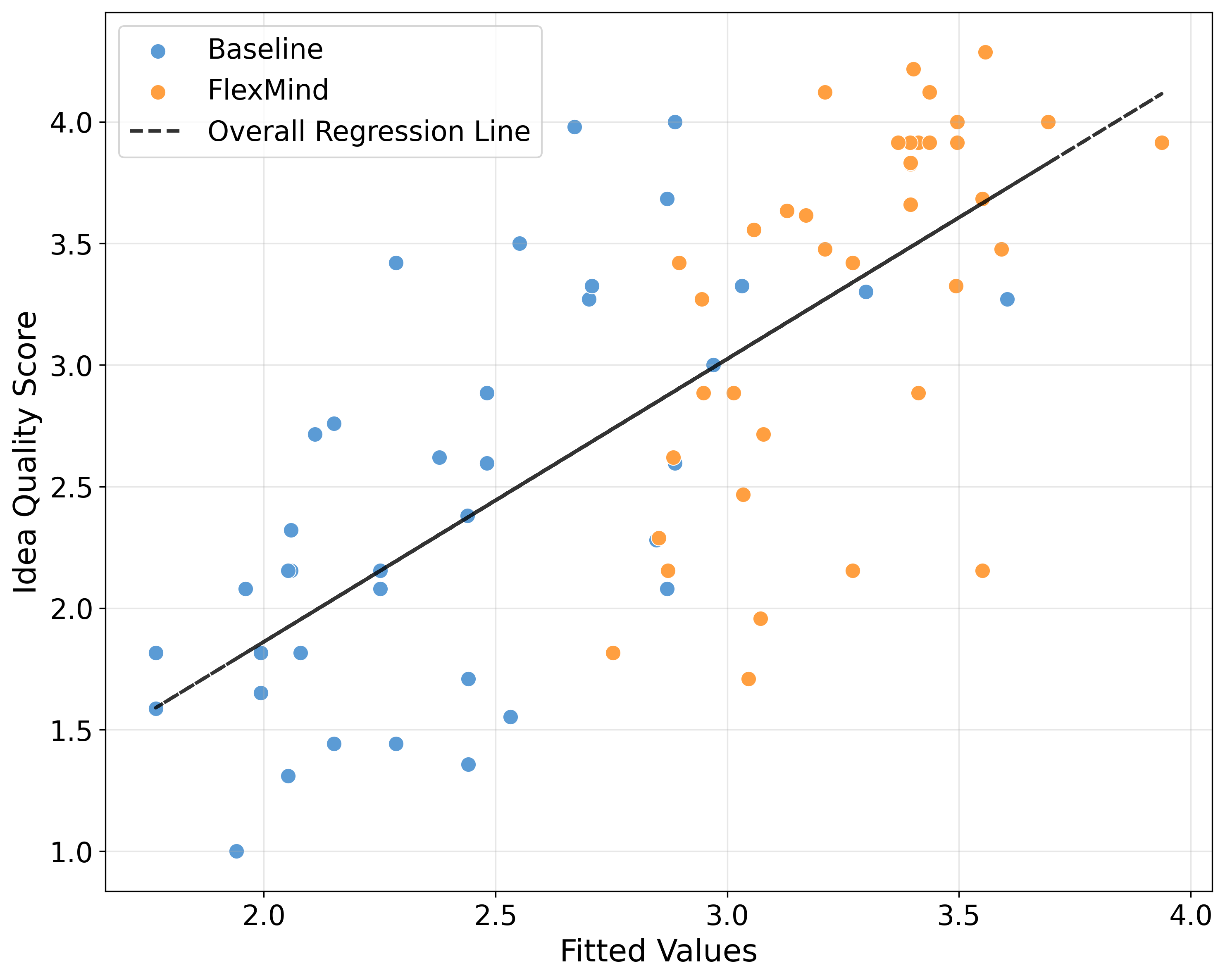}
  \caption{\textbf{Predicted vs. observed idea quality scores from the mixed-effects regression (table~\ref{tab:regression})}. The x-axis shows predicted scores and the y-axis shows observed ratings (1–5). The alignment between predictions and observations demonstrates that the model captures systematic variation in idea quality associated with both experimental condition and idea chain length.}
  \label{fig:regression}
\end{figure}

A linear mixed-effects model predicting idea quality from chain length showed that deeper idea chains were associated with significantly higher-quality ideas(see Table~\ref{tab:regression} and Figure~\ref{fig:regression}). Idea chain depth emerged as a significant predictor, even after accounting for condition, with each unit increase in depth associated with a 0.59-point increase in overall score (SE = 0.17, z = 3.46, p < .001). Model fit indices pointed to reasonable explanatory power, with an in-sample RMSE of 0.61 and a pseudo $R^2 = 0.50$, suggesting that the model accounted for about half of the variance in overall scores. These results suggest that longer chains are associated with higher idea quality, and is consistent with the hypothesis that part of FlexMind's benefits may have come through encouraging deeper engagement with ideas and longer resulting idea chains.

\begin{table}[h]
\centering
\small
\renewcommand{\arraystretch}{1.2}
\setlength{\tabcolsep}{6pt}
\begin{tabular}{lrrrrr}
\hline
Predictor & Coefficient & SE & $z$ & $p$ & 95\% CI \\
\hline
Intercept & 1.54 & 0.29 &  5.26 & $<.001$ & [0.97, 2.11] \\
Condition (system vs. baseline) & 1.22 & 0.36 &  3.42 & $<.001$ & [0.52, 1.92] \\
Idea Chain Length & 0.59 & 0.17 &  3.46 & $<.001$ & [0.26, 0.93] \\
Condition $\times$ Idea Chain Length & -0.43 & 0.18 & -2.42 & 0.016 & [-0.78, -0.08] \\
\hline
\end{tabular}
\caption{\textbf{Mixed-effects regression predicting idea quality score} ($R^2 = 0.50$). Fixed effects include \emph{Condition} (FlexMind vs. baseline; baseline = reference), \emph{Idea Chain Length} (path length from the tree root to the idea’s node) and their interaction. Longer chains predict higher score ($\beta=0.59$, SE=0.17, $z=3.46$, $p<.001$); FlexMind ideas score higher than baseline ($\beta=1.22$, SE=0.36, $z=3.42$, $p<.001$).}
\label{tab:regression}
\end{table}

\begin{figure}[t]
  \centering\includegraphics[width=0.4\linewidth]{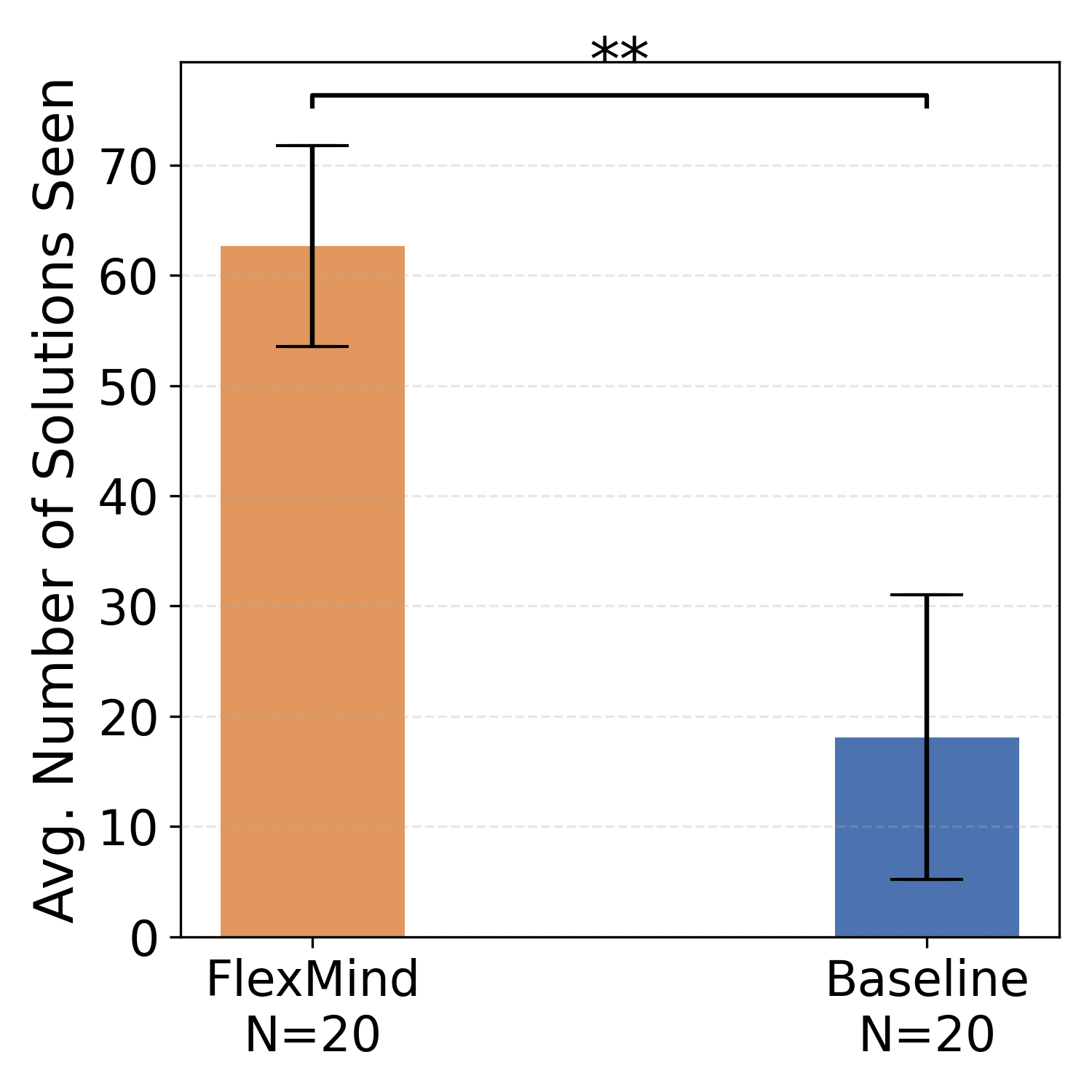}
    \caption{\textbf{Exposure to ideas by condition.} Bars show the average number of solutions participants viewed during ideation, including system-generated and participants’ own written ideas; error bars show standard errors. Participants using FlexMind viewed significantly more solutions than baseline(paired two-tailed t-test, t(19)=14.77, p<.001).}  
    \Description{Bar chart comparing conditions. Participants using FlexMind (N=20) viewed many more solutions on average than baseline participants (N=20). Error bars show standard errors, and a bracket with “**” indicates a statistically significant difference (p < .01).}
  \label{fig:solution_actions_comparison}
\end{figure}

\begin{figure}[t]
  \centering
  \includegraphics[width=0.6\linewidth]{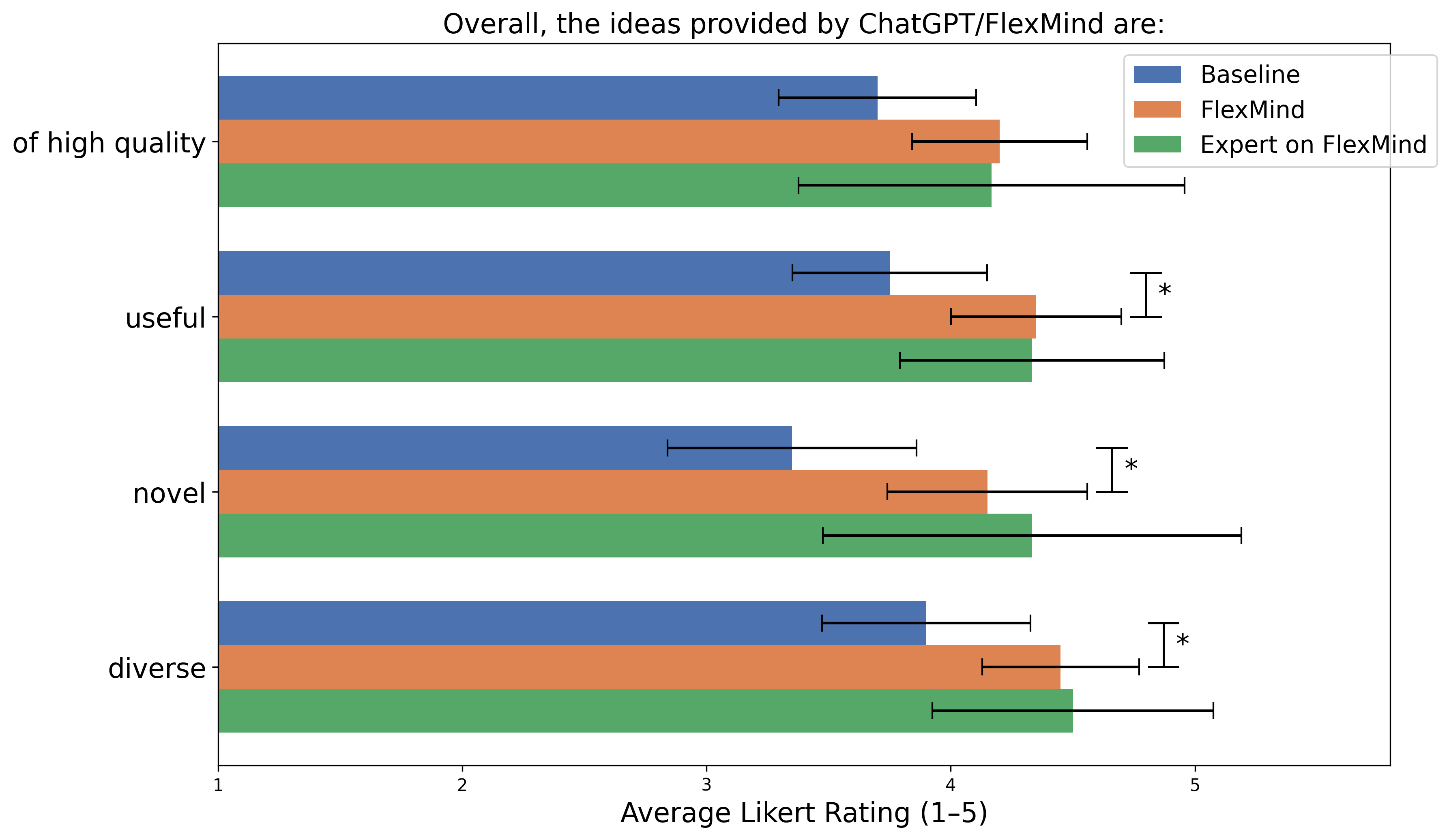}
  \caption{\textbf{User reported ratings (1–5) for provided idea  across four dimensions (quality, useful, novel, diverse).} Blue = Baseline; Orange = FlexMind; Green = Expert case study, where a separate group of experts used FlexMind exclusively. Error bars represent standard errors. Asterisks (*) mark significant differences after Bonferroni correction for Welch’s t-test. FlexMind is overall rated higher than Baseline, with significant gains in usefulness, novelty, and diversity.}
  \Description{Horizontal bar chart with three colored bars for each dimension of idea quality: high quality, useful, novel, and diverse. Blue bars represent the Baseline condition; orange bars represent the within-subjects FlexMind condition; green bars represent an Expert-only group that used FlexMind. The x-axis is average Likert rating from 1 to 5. Error bars show standard errors. Across all four dimensions, the orange bars are longer than the blue bars, indicating higher participant ratings for FlexMind than Baseline. Asterisks mark significant differences for usefulness, novelty, and diversity. The green expert bars are high and generally similar to or above the orange participant FlexMind bars.}
  \label{fig:survey_idea_quality}
\end{figure}

\begin{figure}[t]
  \centering
  \includegraphics[width=\linewidth]{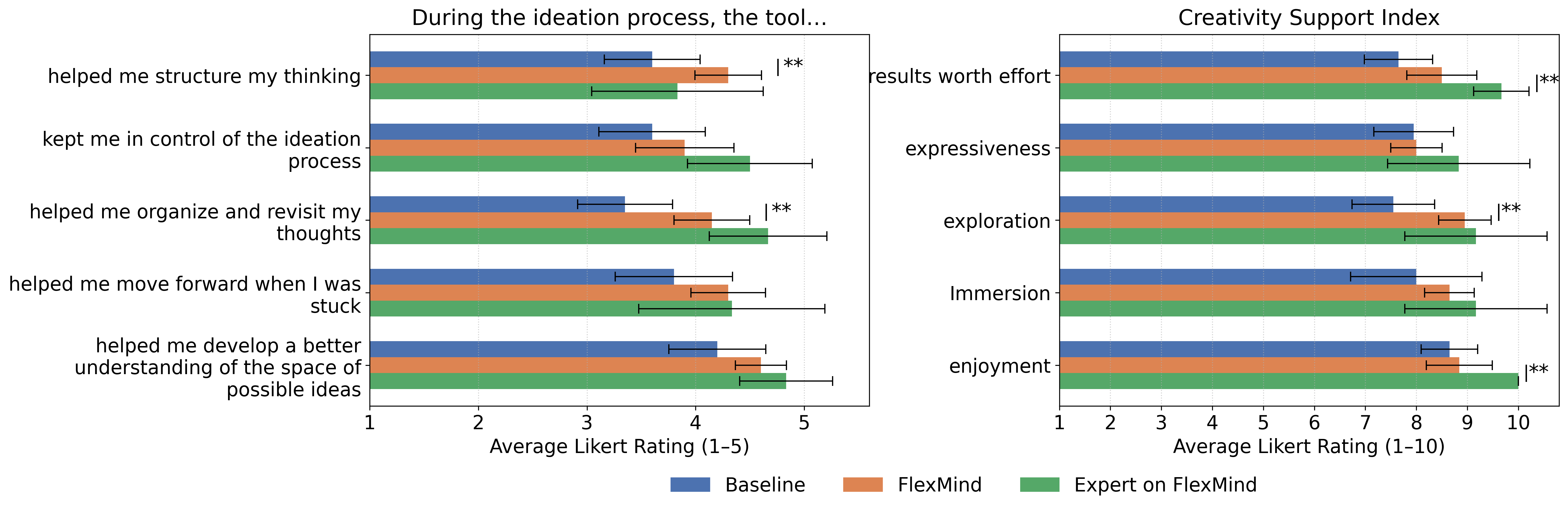}
  \caption{\textbf{Participants’ ratings of ideation support.} Left: Average Likert ratings (1–5) of how well the tool supported participants during ideation. FlexMind was rated significantly higher than Baseline for structuring thinking and organizing/revisiting ideas. Right: Creativity Support Index (CSI) ratings (1–10), excluding the Collaboration factor. FlexMind received significantly higher scores than Baseline in results worth effort, exploration, and enjoyment, withratings from expert case study on FlexMind also high.}
  \Description{Two side-by-side horizontal bar charts. Left panel (Ideation support, 1–5 scale): For five statements—helped structure my thinking, kept me in control, helped me organize/revisit thoughts, helped me move forward when stuck, helped me understand the space of ideas—three bars appear per row: blue (Baseline), orange (FlexMind), green (Expert-only FlexMind). Orange and green bars are generally longer than blue. Significant differences are marked with asterisks for “structured my thinking” and “organize/revisit thoughts,” where FlexMind exceeds Baseline. Error bars show standard errors. Right panel (Creativity Support Index, 1–10 scale, Collaboration removed): Rows for results worth effort, expressiveness, exploration, immersion, and enjoyment, each with blue (Baseline), orange (FlexMind), and green (Expert-only FlexMind) bars. FlexMind (orange) is longer than Baseline (blue) on multiple dimensions—especially results worth effort, exploration, and enjoyment—each marked with asterisks. Expert-only FlexMind (green) bars are also high. Error bars show standard errors.}
  \label{fig:CSI}
\end{figure}

\begin{table}[!ht]
    \centering
    \begin{tabular}{llll}
    \hline
        & Baseline Mean(SD) & \sys Mean(SD) & W (Bonf. p) \\ \hline
        Tree Count & 5.95 (3.30) & 5.55 (2.96) & 83.0 (1.000000) \\
        Nodes Count& 21.90 (11.76) & 43.55 (14.72) & 8.0 ($<$0.01)\textsuperscript{*} \\
        Avg. Tree Depth & 1.59 (0.84) & 2.37 (1.25) & 27.0 ($<$0.05)\textsuperscript{*} \\
        Branch Count & 21.90 (9.26) & 38.00 (13.71) & 8.5 ($<$0.01)\textsuperscript{*} \\
        Avg. Branch Length & 2.58 (0.88) & 3.07 (0.73) & 40.0 (0.108932) \\
    \hline
\end{tabular}  
    \caption{\textbf{Tree-structure metrics by condition.} Values are mean per participant (SD). Compared with Baseline, \sys yielded more nodes (43.55 vs.\ 21.90) and more branches (38.00 vs.\ 21.90), with deeper trees (Avg.\ Tree Depth 2.37 vs.\ 1.59). Tree count and avg.\ branch length did not differ significantly. Tests are paired Wilcoxon signed-rank; table reports $W$ and Bonferroni-corrected $p$; * indicates significance after correction.}

    \label{table:tree}
\end{table}

\subsubsection{\tradeoffmitigation and Exploration Depth}
\label{controlled study result: tradeoff}
The \tradeoffmitigation chain played a central role in the increased depth observed in \sys. Trade-off actions were the most frequent action type (Table~\ref{table: action_type}) and received high usefulness ratings (Table~\ref{table:feature-usefulness}). Participants frequently noted that the trade-offs pushed them to \pquote{question feasibility} (P4) and attend to \pquote{issues I had not considered} (P20). This aligns with quantitative results: the same participants generated substantially more trade-offs in \sys (M=5.60, SD=2.39) than in the baseline (M=0.50, SD=1.05; Table~\ref{table: action_type}) and were more likely to continue along an existing branch when using \sys (M=34.82\%, SD=21.89\%) than when using the baseline (M=24.00\%, SD=14.13\%; Table~\ref{table:jumpdata}). In the baseline condition, trade-offs were rarely surfaced, leading participants to rely on familiar evaluation criteria or switch solutions prematurely. Several participants described the baseline as offering few \pquote{meaningful risks} (P16) or producing ideas that felt \pquote{hard to implement} (P19). Together, these results show how FlexMind’s trade-offs and mitigation suggestions sustained deeper exploration.

\subsubsection{Multi-threaded Structure and Exploration Depth}
FlexMind’s multi-threaded structure also contributed to deeper exploration by helping participants stay within and return to existing branches. Compared to the baseline, participants in \sys were significantly more likely to continue, parallel, or cross within a tree (39.91\% vs. 65.41\% new-tree jumps; W = 7, Bonf. p = .002, see Table~\ref{table:jumpdata}). Participants described the tree view as supporting \pquote{easy comparison} (P8) and helping them \pquote{revisit earlier points} (P11), which is consistent with these behavioral patterns. Importantly, this increase in within-tree development did not reduce the total number of trees explored (Table~\ref{table:tree}). Instead, FlexMind shifted participants away from rapid branch turnover toward sustained work within multiple concurrent threads, enabling deeper refinement without sacrificing breadth.

\begin{table}[!ht]
    \centering
    \begin{tabular}{lccc}
    \hline
        Jump Type & \sys Mean(SD)$\downarrow$ & Baseline Mean(SD) & W (Bonf. p) \\ \hline
        New Tree (pct) & 39.91 (14.58) & 63.25 (18.14) & 3.0 (0.001)\textsuperscript{*} \\
        Continue Branch (pct) & 34.82 (21.89) & 24.00 (14.13) & 55.0 (0.537) \\
        Switch Tree (pct) & 9.80 (11.02) & 0.71 (3.20) & 3.0 (0.038)\textsuperscript{*} \\
        Parallel Branch (pct) & 8.00 (5.06) & 10.64 (13.56) & 60.0 (1.000) \\ 
        Cross Branch (pct) & 7.47 (5.97) & 1.40 (3.54) & 1.0 (0.009)\textsuperscript{*} \\
    \hline
    \end{tabular}
    \caption{\textbf{Tree navigation moves during ideation by condition (mean \% per participant; SD in parentheses). } \emph{New tree}: start a new tree; \emph{Continue branch}: keep working along the same branch; \emph{Switch tree}: move to a different existing tree; \emph{Parallel branch}: start a sibling branch from the same parent; \emph{Cross-branch}: jump to a non-sibling branch within the same tree. \sys shows fewer new-thread starts (39.91\% vs.\ 63.25\%) and more within-tree navigation, especially switches between existing threads (9.80\% vs.\ 0.71\%) and cross-branch jumps (7.47\% vs.\ 1.40\%); differences for continue-branch and parallel-branch are not significant. Tests are paired Wilcoxon signed-rank; table reports $W$ and Bonferroni-corrected $p$; * indicates significance after correction}
    \label{table:jumpdata}
\end{table}

\begin{table}[!ht]
\centering
\small
\renewcommand{\arraystretch}{1.02} % near-default line spacing
\setlength{\tabcolsep}{6pt}
\begin{tabular}{p{0.42\linewidth} p{0.29\linewidth} p{0.29\linewidth}}
\hline
\textbf{Feature Name} & \textbf{Usefulness Rating in the pre-define condition (M $\downarrow$, SD)} & \textbf{Usefulness rating in expert condition (M, SD)} \\
\hline
Externalized thinking steps/flows        & \usecell{4.50}{0.69}{freq_Externalized_thinking_steps_flows.png} & \usecell{5.00}{0.00}{freq_exp_Externalized_thinking_steps_flows.png} \\
Tradeoff analysis                        & \usecell{4.45}{0.76}{freq_Tradeoff_analysis.png}                  & \usecell{4.67}{0.82}{freq_exp_Tradeoff_analysis.png} \\
Adding your own tradeoffs and solutions  & \usecell{4.30}{0.80}{freq_Adding_your_own_tradeoffs_and_solutions.png} & \usecell{4.67}{0.52}{freq_exp_Adding_your_own_tradeoffs_and_solutions.png} \\
Initial overview page                    & \usecell{4.15}{0.67}{freq_Initial_overview_page.png}              & \usecell{3.50}{1.05}{freq_exp_Initial_overview_page.png} \\
Save function                            & \usecell{4.10}{1.07}{freq_Save_function.png}                      & \usecell{4.33}{1.03}{freq_exp_Save_function.png} \\
Solutions for mitigating tradeoffs       & \usecell{4.05}{0.76}{freq_Solutions_for_mitigating_tradeoffs.png} & \usecell{4.17}{0.98}{freq_exp_Solutions_for_mitigating_tradeoffs.png} \\
Question and answer                      & \usecell{3.85}{1.18}{freq_Question_and_answer.png}                & \usecell{4.17}{1.33}{freq_exp_Question_and_answer.png} \\
Similar idea exploration                 & \usecell{3.70}{0.86}{freq_Similar_idea_exploration.png}           & \usecell{4.33}{0.52}{freq_exp_Similar_idea_exploration.png} \\
\hline
\end{tabular}
% \caption{Usefulness ratings for \sys active features from participants in controlled condition and expert condition, measured on a 5-point Likert scale (higher is better).}
\caption{\textbf{Usefulness ratings for FlexMind features in the controlled condition (Baseline vs.\ \sys) and in the expert-only condition.} Ratings use a 5-point Likert scale (higher = more useful) and are reported as mean (SD). Across both groups, participants rated externalizing thinking steps/flows, tradeoff analysis, and adding their own tradeoffs/solutions as most useful, while features such as similar idea exploration and question–answer support were rated somewhat lower.}
\label{table:feature-usefulness}
\end{table}

\subsection{Broader Exploration}
\sys enabled significantly broader exploration during ideation. Participants encountered far more solutions in \sys (M = 62.7, SD = 9.11) than in the baseline (M = 18.1, SD = 12.90, Figure~\ref{fig:solution_actions_comparison}). They also produced nearly twice as many information nodes and more total thinking threads (baseline: $M = 21.9$, $SD = 9.26$; \sys: $M = 38.00$, $SD = 13.71$; Bonferroni-corrected $p < .01$, Table~\ref{table:tree}). Although baseline participants could request more ideas, this was not observed in practice, instead they spent time reviewing the ideas generated in each turn to identify promising ones and crafting new prompts to elicit higher-quality responses. In contrast, \sys continuously surfaced new ideas, matching participant self-reports that the system provided more variety and a broader view of the design space(see Figure~\ref{fig:survey_idea_quality}). These results demonstrate that \sys achieved its design goal of supporting breadth through a multi-threaded structure. The system exposed users to more ideas without reducing their ability to evaluate or elaborate them, complementing the depth effects reported earlier.

\subsubsection{Using Abstraction Helps Expand and Organize Ideas but Also Introduces Constraints}
\begin{figure}[t]
  \centering
  \includegraphics[width=0.9\linewidth]{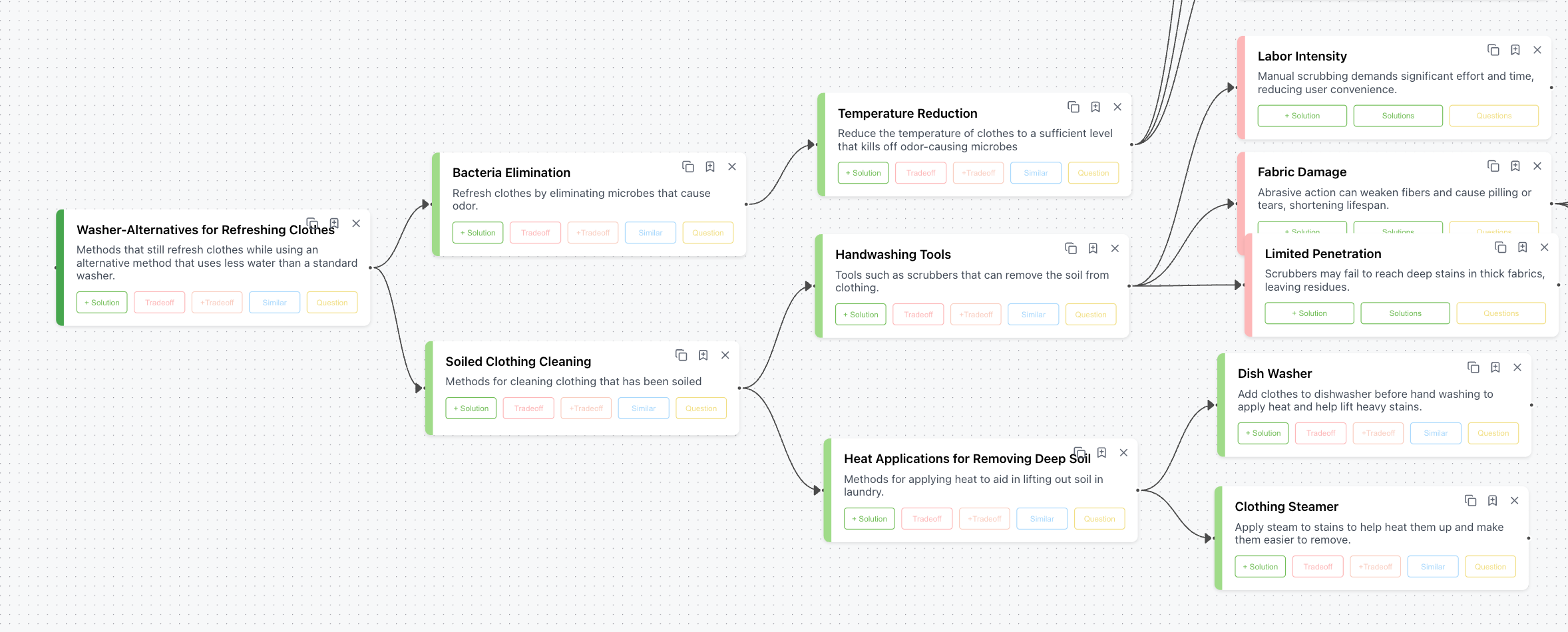}
  \caption{\textbf{Participant-Created Canvas on  FlexMind.} Participant P20 used the system to design alternatives to machine washing for refreshing clothes. Starting from the high-level problem (“Washer Alternatives for Refreshing Clothes”), they broke down possible solution paths, such as eliminating bacteria, reducing temperature, or using solid cleaning methods. They expanded these into concrete workarounds (e.g., handwashing tools, dishwashers) while also documenting trade-offs like labor intensity and limited penetration. This illustrates how participants leveraged the hierarchical canvas to both generate and evaluate workarounds.}
  \Description{Screenshot of a FlexMind canvas with a dotted grid background showing a left-to-right node-link map. A large green card on the left titled “Washer Alternatives for Refreshing Clothes” has arrows to two green child cards: “Bacteria Elimination” and “Solid Clothing Cleaning.” “Bacteria Elimination” branches to “Temperature Reduction” and “Handwashing Tools,” which further connect to “Heat Applications for Removing Deep Soil,” “Dish Washer,” and “Clothing Steamer.” On the far right are three red cards labeled “Labor Intensity,” “Fabric Damage,” and “Limited Penetration,” each receiving incoming arrows from multiple solution nodes. Cards contain short subtitles and small pill labels; arrows indicate parent-to-child relationships. Green cards denote solutions; red cards denote trade-offs.}
  \label{fig:work_around}
\end{figure}

FlexMind uses abstraction to support idea expansion along the \tradeoffmitigation chain, beginning with the initial overview page that groups ideas at a higher level. Most participants found this overview helpful for understanding the overall design space and situating their ideas, noting that it reduced overload and clarified \pquote{where everything fits} (P1). However, abstraction was not uniformly helpful. Participants gave lower ratings (see Table~\ref{table:feature-usefulness}) for the 'Similar' feature which generates additional ideas by abstracting from the current one. They reported that when using the 'Similar' button, some abstracted concepts were difficult to relate back to the original idea, leaving them without a clear conceptual anchor for further ideation(P15, P20, P18, P4, P3, P16). To address this, some participants developed a workaround on the canvas page. They manually added more specific idea nodes beneath a broader category to create structures that better reflected their own interpretation of the idea (see Figure~\ref{fig:work_around}). This helped them restore grounding while still benefiting from the ability to expand ideas.

The findings suggest that accurate abstraction can support broader exploration when users can understand them well, and overly generic or mismatched abstractions can hinder expansion along the \tradeoffmitigation chain. In the expert case study (Section~\ref{expert_section}), we further show that topic familiarity makes a difference: expert participants have made productive use of abstraction for finding new ideas.

% \subsection{Active Thinking and Agency}

% To understand how \sys influenced the creative process itself, we analyzed participants' interactive behaviors to assess their level of active engagement and agency. This analysis focused on quantitative measures of user actions and subjective ratings of the creative experience, revealing a shift towards more reflective and evaluative behaviors in the \sys condition.

\subsubsection{Multi-threaded Structure and Exploration Breadth}
FlexMind’s tree view also broadened exploration by supporting tracking ideas. Participants rated \sys higher on “organizing and revisiting thoughts” reported the tree helped them manage multiple threads and  return to earlier ideas. Users described it as \pquote{efficient for tracking ideas} (P4), helpful for \pquote{seeing how I progressed} (P17), and aligned with their natural grouping strategies (P5). This manifested as more branches and more within-tree expansion in \sys (M=38.0, SD=13.7, Table~\ref{table:tree}) than baseline(M=21.9, SD=9.3, Table~\ref{table:tree}), showing that the multi-threaded structure facilitated wide yet navigable exploration.

\subsubsection{\tradeoffmitigation and Exploration Breadth}
Beyond deepening reflection, participants reported that the \tradeoffmitigation thinking pattern also helped them expand their exploration space. Participants reported that evaluating limitations led them to new directions, even when some trade-offs felt \pquote{impossible} (P2) or when solving a limitation revealed \pquote{ideas I did not think of initially} (P16). These insights are consistent with the observed increase in solution diversity: the \tradeoffmitigation chain generated new pivots that expanded the space of alternatives, complementing the structural breadth afforded by the tree. The following expert case study further validates that the \tradeoffmitigation structure supports the expansion of the solution space.

% \subsubsection{Customized Q\&A}
% Besides tradeoff mitigation, we also observed that participants used the Q\&A feature to seek higher-level analytical information, and they appreciated having the flexibility it provided. For example, P3 mentioned, \pquote{I feel like the question—the one where I could ask the system a question... so we do have a back-and-forth conversation... that makes me feel like I am controlling the process of iterating the design solutions instead of being led by AI.} Similarly, P2 asked questions like, \pquote{What are the estimated costs per load?}, during the process, later noting that \pquote{the ability to directly query very specific information I proposed made the process more efficient. }

\section{Validation Case Study With Expert Practitioners}
\label{expert_section}
We complement the controlled study with a case study to examine how experts used \sys when working on self-chosen problems within their own domains. While controlled tasks allow for precise benchmarking, professional practice is constrained by deep domain familiarity, specialized terminology, and complex trade-offs. 
% While participants in the controlled study had professional engineering or design experiences, they worked on assigned tasks outside their specific expertise. In contrast, real practice is affected by domain familiarity, including concepts, constraints, and terminologies, which influence how people interpret abstractions, reason about trade-offs, and navigate idea spaces. 
This study examines how \sys functions within these authentic workflows. Rather than comparing against a baseline, the expert study therefore examines \sys under conditions that more closely reflect professional workflows.

\subsection{Study Method}

\begin{table}[h]
\label{table:expert_task_des}
\centering
\small
\renewcommand{\arraystretch}{1.3}
\begin{tabularx}{\linewidth}{cX}
\hline
\textbf{ID} & \textbf{Design Task in Expert Condition} \\
\hline
1 & Tools that compensate for hand tremors in fine work. \\
2 & When you’re at sea at night, it’s important to recognize navigation lights of other ships. This is difficult to practice except at sea because the difficulty comes from the movement of the boats in the waves, fog/haze, the relative direction of the boats, and everything is in motion. Also, you have to memorize the light configurations (tug, fishing boat, etc.). How to learn and practice this? \\
3 & Design a way for autonomous cars to reliably transfer perception data to each other as they drive in an environment with human passengers. \\
4 & Golf cart that drives itself and follows the golfer naturally. \\
5 & Recognizing hand gestures from continuous hand interaction in live presentation settings. \\
6 & Design a system that automatically adjusts a car seat to the driver’s most comfortable position, eliminating the need for manual adjustment. \\
\hline
\end{tabularx}
\caption{Design tasks used in expert condition}
\label{table:design_tasks_expert}
\end{table}

\subsubsection{Participants \& Procedure} 
Recruiting true R\&D experts is non-trivial; we leveraged professional networks and a targeted mailing list to secure volunteers with 4–10+ years of domain practice across design, sensing/hardware, robotics, and industrial research. We recruited six experts with substantial design/engineering experience: 1) a senior designer (>10 years), 2) a sensing/hardware research scientist (>4 years in a research department), 3) a robotics systems engineer (>10 years in industry), 4) an industrial researcher in design (>10 years), and 5) two senior PhD researchers (fabrication in HCI; sensing) with design backgrounds (each >8 years combined design/engineering experience). All experts volunteered to participate in an one-hour session. 

The procedure mirrored the controlled study: remote session (60 minutes; Zoom), screen-share, think-aloud, standardized recording of “keepable” ideas, and a short post-session survey. As in the controlled \sys condition, experts first watched the same tutorial video and completed the same tutorial task to learn \sys before using it on their main tasks. Crucially, participants worked on \textbf{self-chosen design problems} from their own domains (Table~\ref{table:design_tasks_expert}) that they had not yet fully solved. This ensured that the constraints and criteria used to evaluate \sys were grounded in real-world professional requirements rather than artificial task prompts.

\subsubsection{Measurement and Analysis}
We collected (i) \sys interaction logs, which automatically capture actions and information as tree structures, and (ii) post-session questionnaires identical to those used in the controlled study (Section~\ref{controlled_study_method}). Expert-condition ratings are reported alongside controlled-condition ratings in the Results (Figure~\ref{fig:survey_idea_quality}, Figure~\ref{fig:CSI}, and Table~\ref{table:feature-usefulness}). Analysis focused on inductive thematic coding of think-aloud and exit interviews and descriptive analysis of survey ratings, following the same procedures as the controlled \sys condition.

\subsection{Findings}
Overall, experts gave high ratings to the system's utility, specifically valuing the \textbf{Trade-off} ($M=4.67, SD=0.82$) and \textbf{Mitigation} ($M=4.17, SD=0.98$) features (see Table~\ref{table:feature-usefulness}). Qualitative analysis showed that experts found the system particularly valuable for supporting three key cognitive activities involved in their exploration of the solution space: reducing ideation stalls by surfacing knowledge experts were less familiar with, overcoming working memory limits through the multi-threaded \tradeoffmitigation structure, and using abstraction to break fixation and clarify vague concepts in order to develop new ideas.

% experts used the system to expand the space through reflective engagement with ideas, flexible movement across parallel threads, and by leaning on abstraction to connect concepts. Taken together, these patterns suggest that in real professional practice, \sys supports not incremental refinement, but transformative ideation: reflection pushed breadth, parallel navigation enabled reframing, and abstraction surfaced unexpected conceptual relationships in the space.

% \subsubsection{ \textbf{Establishing the Chain: Transforming Trade-offs into Continuous Iteration.}}
\subsubsection{{\textbf{Reducing Ideation Stalls by providing knowledge experts are less familiar with}}}

We observed that during ideation, experts can become stalled when progress depends on knowledge they are less familiar with. This reflects a known challenge in professional design, where even experts do not have deep familiarity with every component of a complex system~\cite{cross2004expertise}. When such gaps arise, it's hard for people  to identify concrete trade-offs and possible mitigations that ground further reasoning, thus becoming stalled. While prior work argues that LLM-generated ideas are often too generic to support expert reasoning~\cite{fukumura2025can, collier2025good}, our results suggest that \sys supports experts by surfacing trade-offs and mitigations they are less familiar with, yet can still readily understand because the knowledge remains within their general domain. The generated \tradeoffmitigation chain supports experts to resume analysis and continue developing their ideas, rather than remaining stalled.

% Trade-offs: what issues did they have with not knowing the trade-offs, what value did knowing the trade-offs provide (determining feasibility, etc.), how flexmind helped suggest trade-offs and that experts found that useful.
% Mitigations: how was not knowing mitigations stopping people prematurely, what kinds of things did flexmind propose and how would the users know if those mitigations were actually viable or promising versus bad ones.
% [the below one points out the underlying reason, but not the actual problem/issue we are dealing with]

% Experts found the trade-offs useful because the analysis was relatively complete and covered areas outside their direct expertise.

\begin{figure}[t]
  \centering
  \includegraphics[width=\linewidth]{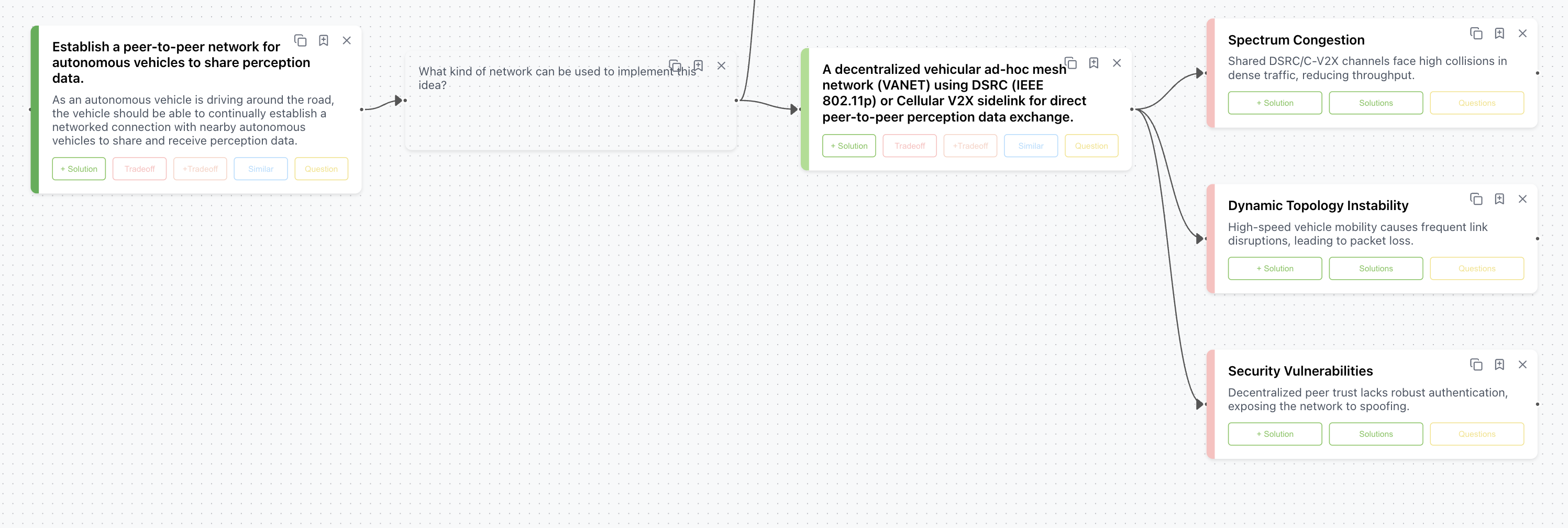}
\caption{In the task ``Design a way for autonomous cars to transfer perception data to each other'', Expert 3, a sensing specialist, used \sys’s \tradeoffmitigation chain to evaluate a less familiar network approach. The system surfaced concrete trade-offs, helping the expert to assess the idea using their general engineering expertise, and avoiding stalling at the evaluation.}
  \label{fig:expert_3}
\end{figure}

\textbf{Experts leveraged system-generated trade-offs to evaluate ideas that involved knowledge they were less familiar with.}
For example, Expert 3, a sensing specialist, who explored using networks to transfer perception data among autonomous vehicles. While the expert was confident in sensing and data acquisition, they found it difficult to evaluate network options due to lower familiarity with network technologies. As a result, they became stuck when attempting to identify concrete trade-offs of using a decentralized vehicular ad-hoc network, where vehicles communicate directly without fixed infrastructure. \sys surfaced trade-offs such as rapid changes in vehicle speed and distance, which causes network links to break, leading to disconnections (Figure~\ref{fig:expert_3}). These trade-offs are critical constraints that the expert had not yet considered. With their general engineering expertise, the expert easily understood the trade-offs and recognized their importance. They further used the identified trade-offs to assess the idea feasibility, concluding that this idea was viable but would require physical testing for improving reliability. They later reflected: \pquote{I'm not super familiar with network technology, but (\sys) was able to automatically compare the disadvantages and benefits. It was really helpful because it created more of a thinking process for me to understand pragmatically how realistic it would be.} By externalizing the concrete trade-offs in areas of lower familiarity, \sys worked in tandem with experts' domain knowledge to unblock stalled reasoning and support experts' assessment of ideas. 
% Across all six expert cases, their evaluation stalled at points where ideas depended on knowledge the experts were less familiar with. By surfacing concrete trade-offs, \sys supported experts to apply their existing domain expertise to evaluate ideas.

\begin{figure}[t]
  \centering
  \includegraphics[width=\linewidth]{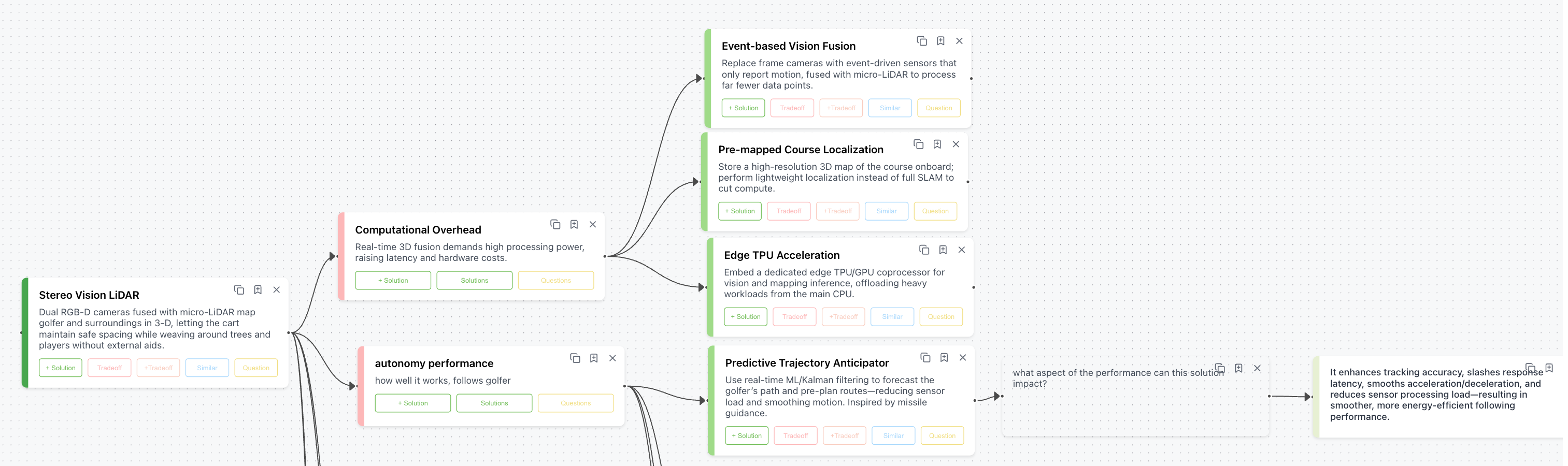}
\caption{In the task ``design a golf cart that drives itself and follows the golfer naturally'', Expert 4 used \sys to examine a camera plus LiDAR tracking approach. After trade-offs such as computational latency and tracking reliability were identified, the generated mitigations introduced concrete follow-up directions, supporting the expert's continued refinement rather than ideation stall.}
  \label{fig:expert_4}
\end{figure}

\textbf{When trade-offs were identified, the system-generated mitigations helped experts to move past reluctance to confront flaws and to transform unresolved trade-offs into concrete solutions as next steps; otherwise, they might have stalled at identified issues.}
% \textbf{\sys reduced this burden by surfacing concrete paths forward that lowered the reluctance to confront flaws, enabling continued progress rather than stalling at unresolved trade-offs.}
Consistent with the controlled study, experts found it cognitively demanding to generate feasible mitigations once trade-offs were identified. Receiving paired mitigations reduced this pressure, allowing experts to continue reasoning without worrying about generating solutions on their own. Experts described trade-offs as moments where excitement about an idea conflicted with uncertainty about how to resolve its weaknesses. As Expert 1 explained, \pquote{I know there is a trade-off, but because I am too excited about the idea, I am reluctant to think about that... If I already know some trade-off, I would be very worried because I cannot think of a quick solution... This pipeline helped me be more confident about the idea and try to think through the whole idea without me putting a lot of effort in other aspects}. 
In addition, experts also recognized that the mitigations generated provided concrete starting points to improve ideas when trade-offs emerged. 
For example, when working on a golf cart that automatically follows the golfer, Expert 4 was initially excited about combining stereo cameras with micro-LiDAR detect the user's location (see Figure~\ref{fig:expert_4}) but hesitated to proceed after identifying major trade-offs such as computational overhead which causes latency, and potential low accuracy of tracking. However, \sys ’s mitigations, such as leveraging ML models to predict the golfer's path and reducing sensor load, introduced directions the expert had not initially considered and inspired them to continue developing the concept rather than abandoning it. Expert 4 described  suggestions as \pquote{very useful} and \pquote{speed me up in that process.} 
% They explained that receiving these early mitigations allowed them to identify missing pieces during prototyping: \pquote{in the very beginning phases when you are prototyping, this helps you fill the gaps like: Oh, what else have I not thought of?} The paired mitigation scaffold reduced this burden by transforming critique into a forward-moving process rather than a terminal judgment.
By pairing trade-offs with concrete, system-generated mitigations, the system reduced reluctance to confront flaws and transformed tradeoffs into forward-moving solutions, supporting sustained elaboration of promising ideas instead of premature abandonment.

\subsubsection{\textbf{
Overcoming Working Memory Limits Through Multi-Threaded \tradeoffmitigation Chains for Comparing and Managing Idea Alternatives
% Networking the Chains: Comparing and Transferring Reasoning Across Ideas
}}
Beyond supporting ideation depth within a single branch, organizing \tradeoffmitigation chains in a multi-threaded format addressed a key challenge faced by experts: limited working memory~\cite{cowan2012working}. We observed that experts did not treat ideas in isolation. Instead, they frequently compared alternatives, with each idea involving multiple trade-offs that needed to be weighed against others. In addition, new inspirations could arise at any point, including while experts were working within an existing branch, requiring them to quickly shift attention to another idea. Both behaviors can easily exceed working memory without external support. The multi-threaded \tradeoffmitigation structure in \sys provided an organized representation of ideas and tradeoffs, along with a flexible way to add and manage idea branches. This supported experts’ need to compare alternatives, and branch at any moment to capture new inspirations while preserving prior ones. These findings extend the controlled study results, where some participants noted that the multi-threaded structure helped them compare or keep track of ideas. For experts in the session, who were more familiar with the domain and likely had more thoughts on the task, their comparison of ideas was not occasional but continuous, and they frequently moved between branches to develop ideas as new thoughts emerged. The results showed that the multi-threaded structure supported this form of sustained analysis under higher cognitive load.

\begin{figure}[t]
  \centering
  \includegraphics[width=\linewidth]{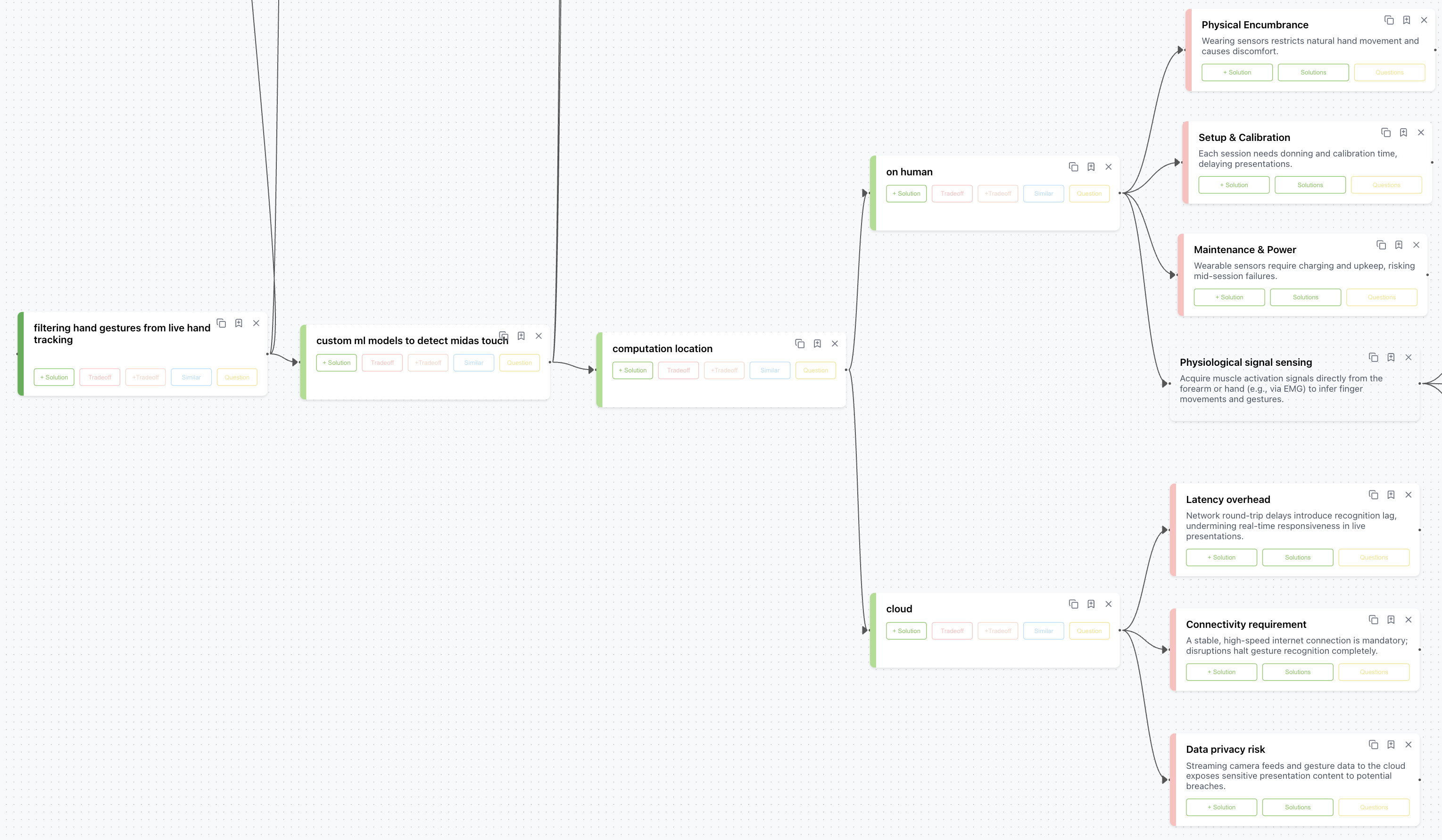}
\caption{In the task ``recognizing hand gestures in live presentation settings'', Expert 5 used \sys’s multi-threaded structure to place two parallel solution approaches: on-device and cloud-based data processing, into a shared view, which enables a direct comparison of multiple constraints.
}
  \label{fig:expert5_across}
\end{figure}

\textbf{Experts compared ideas with multiple trade-offs using parallel branches.}
When exploring complex ideas, experts often need to reason about trade-offs across alternative solutions. This comparison becomes difficult when each idea involves multiple trade-offs that should be considered simultaneously. We observed that the multi-threaded chains supported experts to easily compare ideas by presenting their trade-offs side by side. For example, when designing a hand gesture recognition system for live presentations, Expert 5 added both on-device and cloud-based data processing as two alternatives(Figure~\ref{fig:expert5_across}). However, each idea involved multiple trade-offs, making the comparison hard. To support holistic comparison, the expert rearranged the branches to place them next to each other.
Through this side-by-side view, the expert identified the core tension out of multiple tradeoffs: the on-device approach required bulky hardware that restricted natural hand movement, whereas the cloud-based approach introduced latency issues. Reflecting on this process, Expert 5 noted that having solutions and their trade-offs structured in parallel was unexpectedly helpful: \pquote{it turned out to be a nice thing in the system, which I did not realize, was when you specifically have solutions and trade-offs in that kind of structure (the trade-offs for two ideas listed in parallel branches)
% I might not have looked into the trade offs directly if I was working with ChatGPT. 
... I would eventually come to the point where I would need the trade-offs to compare two different solutions}
% , but upfront I might not have thought of that. Actually, it was nice to have them now.  
By making trade-offs visible in an easy-to-compare format, the multi-threaded structure supported deeper comparison and continued ideation across alternatives.

\begin{figure}[t]
  \centering
  \includegraphics[width=\linewidth]{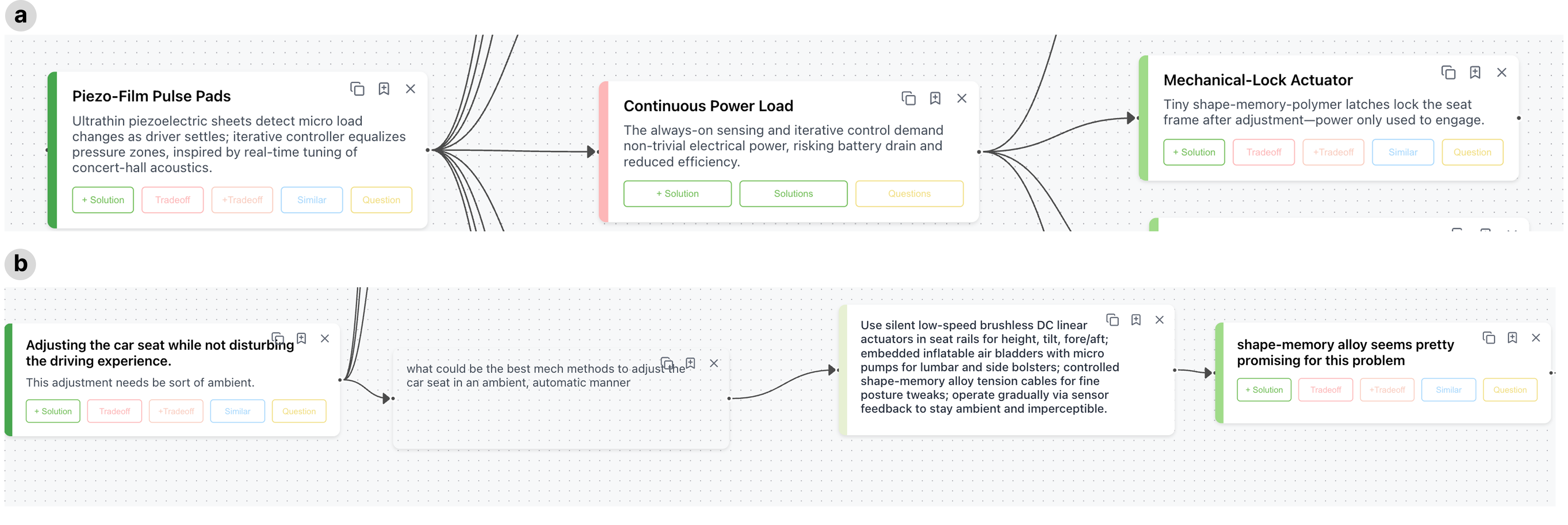}
\caption{In the ``designing adaptive car seats'' task, Expert 6 used \sys’s multi-threaded interface to explore alternative solution paths in parallel. After a power-consumption trade-off was identified in a sensing approach, the system’s mitigation suggestion prompted the expert to pivot into a new branch centered on mechanical adjustments while preserving the original thread. The figure shows how insights from one branch informed the creation of another without discarding prior work.}
  \label{fig:expert_6}
\end{figure}

\textbf{Experts used the multi-threaded structure to flexibly add and manage multiple idea branches.} 
Experts’ inspirations can arise at any point during ideation, creating a challenge in recording, managing, and developing emerging ideas. Rather than staying within a single branch, we observed that experts often carry insights from one idea to another to form new alternatives. This process places a heavy burden on working memory, as experts must preserve ongoing reasoning while simultaneously starting new branches. The multi-threaded structure in \sys addresses this challenge by making it easy to add new branches and move between them, allowing experts to branch at any moment without losing prior work.
For example, Expert 6 initially pursued a sensing-based approach to detect driver position for adjusting a car seat for comfort (Figure~\ref{fig:expert_6}). When the system surfaced the trade-off of continuous power consumption and suggested a mechanically locking component as a mitigation, the expert reconsidered the overall solution. They wanted to explore this new direction while keeping the original sensing-based idea available for later revisiting. Using the multi-threaded structure, the expert created a new branch focused on purely mechanical adjustment strategies while maintaining the sensing-based branch in parallel. 
This pivot led to exploration of alternatives such as inflatable air bladders embedded in the seat and shape-memory alloy tension elements that could adjust posture, which the expert described as \pquote{thought-provoking topics}.
By supporting flexible branching and easy navigation across ideas, the system enabled experts to transfer insights across solutions while preserving earlier ideas and reasoning.

% Expert 6 explained: \pquote{I think doing this changed my mind. I find some thought-provoking topics.} 

% As shown in Figure~\ref{fig:expert_6}, Expert 6 initially fixated on a piezoelectric self-adjustment approach in the adaptive car seat task (Table~\ref{table:design_tasks_expert}). When a power–load trade-off surfaced and a mechanical-lock actuator was suggested in one chain, the expert directly used this insight to initiate a new thread focused on mechanical solutions, enabling exploration of more energy-efficient adjustment mechanisms.This shift in reasoning reframed how they understood the problem and opened new directions. 

% \textbf{Experts also used the multi-thread structure to manage and sustain different ideas.} Figure~\ref{fig:expert_1_across} shows how \sys helped Expert 1 identify several critical trade-offs in their initial idea and develop multiple mitigations in response. The multi-thread \tradeoffmitigation structure allowed the expert to keep promising branches active while freely exploring others. By maintaining these threads simultaneously, the expert could drill down into the specific context of each path. This specificity allowed structurally distinct ideas to emerge and coexist on the canvas, resulting in greater overall diversity than in the initial exploration. As Expert 1 remarked: \pquote{I can see some similarity among the first round of the ideas. When I go to the canvas, I think the ideas are more diverse. Maybe because the topic is more specific.}

% \begin{figure}[t]
%   \centering
%   \includegraphics[width=\linewidth]{parts/figure/expert1_across.png}
% \caption{ \sys helped Expert 1 surface three critical trade-offs and multiple mitigations. The multi-thread \tradeoffmitigation structure ensured a relatively broad design space that people can actively engage with.
% }
%   \label{fig:expert_1_across}
% \end{figure}

\subsubsection{\textbf{Using Abstraction to Break Fixation and Clarify Vague Concepts for Developing New Ideas}}
While the multi-threaded \tradeoffmitigation structure supported idea development, experts sometimes reached points where refining current chains was insufficient. At these moments, we observed that they became fixated in local solution spaces or struggled to articulate early ideas clearly. The expert case study revealed abstraction, which was less useful for novices in the controlled study (Figure~\ref{table:feature-usefulness}), as a key mechanism for extending exploration beyond existing chains. By generating abstract concepts, \sys helped experts break fixation and clarify vague ideas, seeding new idea chains.

% this extends the results from the controlled study, where Results from the controlled study suggested that abstraction-based exploration was difficult for novices to use productively (see Figure~\ref{table:feature-usefulness}). Many reported that the generated abstractions as detached from their original ideas. 

\textbf{Expert used the abstract concepts generated through `similar' feature to surface unexpected components of ideas that triggered pivots into novel, orthogonal solution spaces. } For example, Expert 2, who was exploring VR for sailing training, encountered the idea of using gamification (i.e., scoring the user's speed and accuracy for recognizing various sailing signals) as part of an idea about a VR training platform. Recognizing an interesting new solution space they hadn't previously considered, the expert added it as a pivot point (using the 'similar' feature). They explored new ideas that used gamification in non-VR settings, such as using a phone's gyroscope to simulate ship pitch and roll or head to head competitions between players racing to a target light at night, which they described as \pquote{good ideas}. Another way that experts discovered unexpected yet relevant opportunities was by using the similarity button explicitly. Expert 5 described using the similarity feature as a way to search for things they wouldn't even know to search for: \pquote{When clicking on “Similar,” it came up with a few things, some of which I would have never imagined at all, and I would have never searched for in the first place, because I did not even know those things existed. So it is like a wild card... I start to understand what similar things exist that I might have missed if I were thinking more structurally on my own.}

This pattern of finding unexpected new directions and using them to pivot to new solution spaces was common across experts. However, it was not common in the controlled study where people were doing it for fixed experimental tasks. This may be because controlled study participants didn't have enough domain knowledge to make unknown unknowns as valuable since many of the ideas they were coming across were new to them. For experts working on their own tasks, in contrast, breaking fixation and finding new orthogonal solution spaces appeared to be one of most valuable contributions of the system.

\begin{figure}[t]
  \centering
  \includegraphics[width=\linewidth]{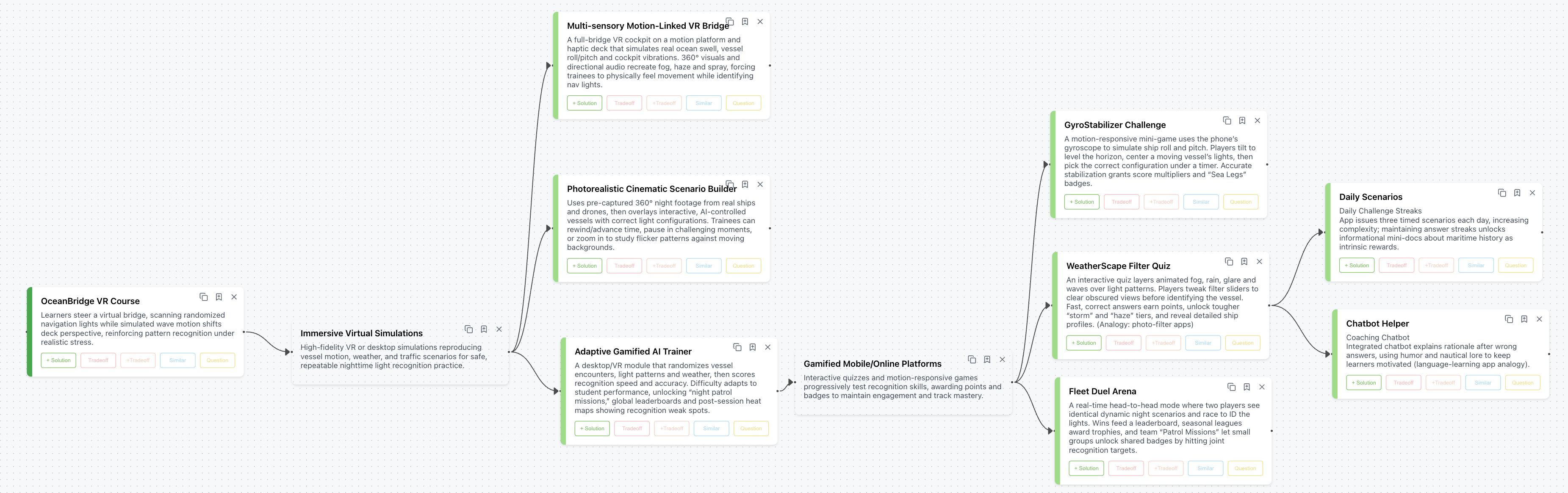}
\caption{ In the ``sailing navigation training'' task, Expert 2 used abstract concepts generated by \sys’s Similar feature to pivot into an orthogonal solution space, gamification, which helps the expert to explore ideas they had not considered before.
}
  \label{fig:expert2_up}
\end{figure}

\begin{figure}[t]
  \centering
  \includegraphics[width=\linewidth]{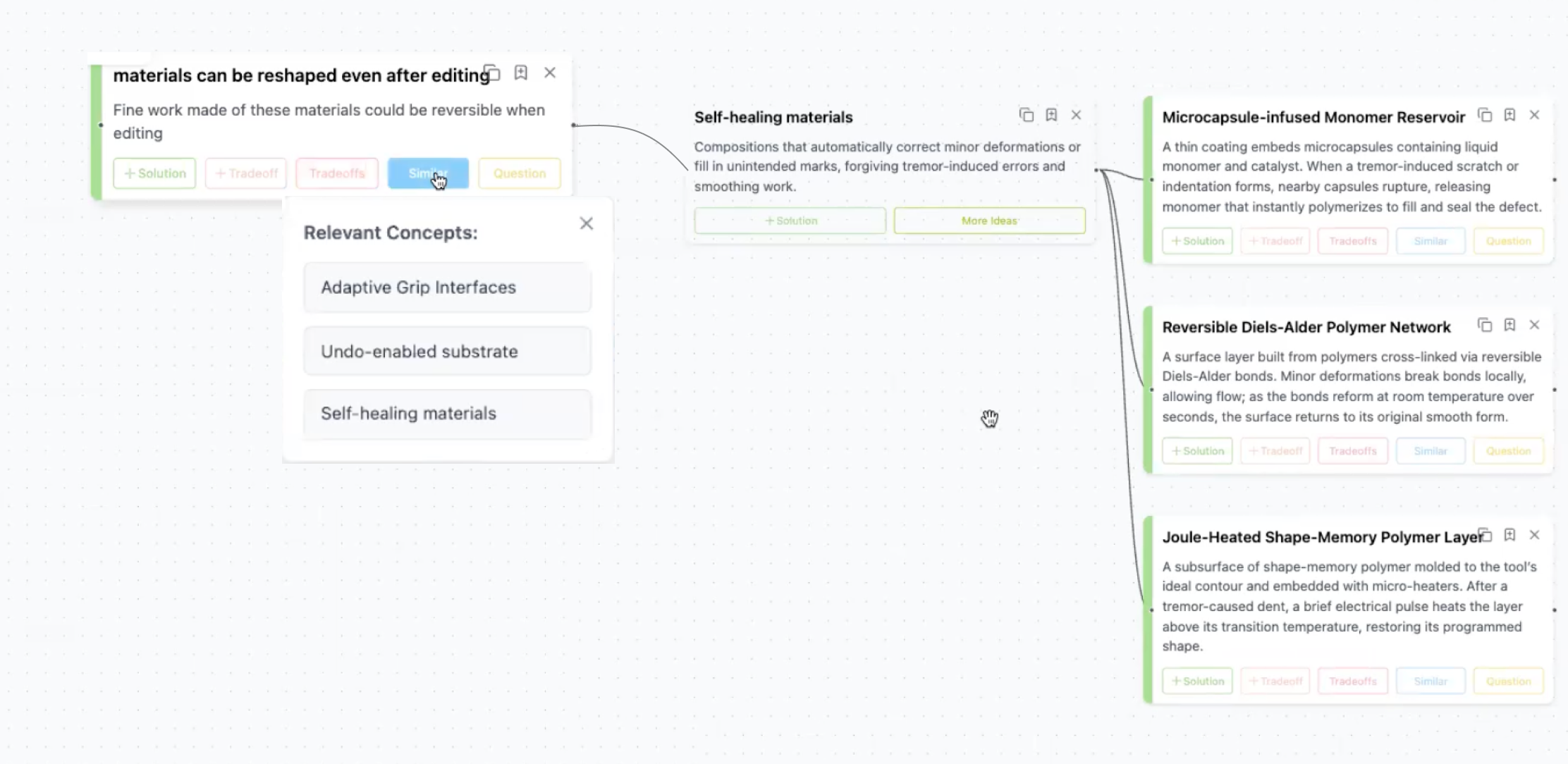}
\caption{ In the task of ``compensating for hand tremor during fine work'', Expert 1 used an abstraction generated by \sys to clarify an initially vague material-based intuition. The system articulated the idea as a self-healing material concept, which the expert then used as a concrete conceptual anchor to explore specific implementation directions.
%, such as coatings that automatically repair small surface defects.
}
  \label{fig:expert_1_up}
\end{figure}

\textbf{Beyond enabling pivots into new solution spaces, the abstractions generated by \sys helped surface and formalize underspecified ideas, transforming vague initial intuitions into explicit concepts that could support productive analysis and further ideation.}  Figure~\ref{fig:expert_1_up} illustrates this process for Expert~1, who was exploring ways to compensate for hand tremor during fine work. The expert initially imagined using materials that could be reshaped after editing so that small tremor-induced errors would not become permanent. While this intuition guided their thinking, they struggled to articulate a precise material concept or to identify defining properties that would support deeper reasoning. The abstract concept generated by \sys, \textit{self-healing materials that automatically smooth minor deformations and fill unintended marks}, made this implicit idea explicit and provided a clearer direction for exploration. The expert used this concept to find specific solutions like: \textit{a surface coating that contains tiny capsules of repair material}. When a tremor causes a small scratch or dent, the capsules break and automatically fill in the mark, preventing minor errors from becoming permanent. Reflecting on this moment, the expert noted, \pquote{Maybe I did not convey this idea very clearly. So I want to explore similar communication of the idea or more specific implementation. For example, the self-healing idea describes the idea better.}

Across three experts in the study, including the example above, even domain experts who know the terminology well often began with ideas that were underspecified and difficult to articulate at their earliest stage. Abstraction in \sys helped turn these early intuitions into explicit concepts, supporting the generation of richer follow-up ideas and further meaningful analysis.

% A similar pattern appeared for Expert 4 in Figure~\ref{fig:expert4_up}, where abstraction around explainable decision rationale helped move beyond interface details and seed several new chains.

% Later analysis based on the vague ideas produces little value because the reasoning is grounded in an imprecise concept. By externalizing and refining this abstraction, the system transformed a loosely defined intention into a clearer conceptual construct, which then seeded a richer set of follow-up ideas. The generated explicit abstractions helped experts articulate the underlying principle they had in mind, turning loosely defined thoughts into precise and generative concepts.

% Taken together, these findings show that abstraction in FlexMind helps experts find related ideas and clarify vague concepts, which support exploration beyond the initial ideas. Abstraction allows insights from one part of the design space to be generalized and used to seed new, coherent chains of reasoning, making it a central driver of conceptual breadth.

\section{Discussion}
\sys was motivated by prior work characterizing creative progress as an iterative process of engaging with constraints, which involves identifying trade-offs in emerging ideas, then reasoning through those trade-offs to develop mitigation strategies~\cite{mumford2002leading, nickel2022manipulating}, and exploring multiple alternatives in parallel as understanding evolves~\cite{dow2010parallel,tohidi2006getting}. 
To support these needs, \sys integrates LLM-generated \tradeoffmitigation chains into a structured, multi-threaded workspace.
% \sys was designed to support the dual needs of ideation: drilling down within an idea through \tradeoffmitigation reasoning~\cite{mumford2002leading, nickel2022manipulating}, and developing multiple idea paths that allow people to compare, shift between, and build across alternatives~\cite{dow2010parallel,tohidi2006getting}.
% To achieve this, \sys pairs LLM-generated \tradeoffmitigation chains with a structured multi-threaded exploration interface. 
We found that this combination supported participants to engage more deeply within individual idea chains while also exploring more broadly across chains, ultimately producing higher-quality ideas. The expert case study further showed that \sys supported professional ideation workflows by reducing ideation stalls through surfacing unfamiliar but relevant knowledge, overcoming working memory limits when managing and comparing multiple ideas, and seeding new idea chains through abstraction when experts needed to break fixation or clarify early concepts. The following discussion reflects on the design implications of these findings for human–AI co-ideation processes, with relevance beyond ideation to broader forms of creativity support using AI.

\subsection{\tradeoffmitigation in Human-AI Co-Ideation and the Prospect of Automation}
Our findings suggest that the \tradeoffmitigation structure significantly contributes to ideation depth, leading to higher-quality ideas. This aligns with prior work suggesting that innovation is driven by the iterative process of identifying trade-offs and resolving them to find new solutions~\cite{ward1999creative,tromp2024creativity}. In the co-ideation process centered on this structure, we observed distinct, complementary roles of AI and humans: the AI surfaces trade-offs and mitigations that users—even domain experts—may not immediately perceive; humans are inspired by the system's output to inject new constraints and ideas, while applying the necessary intuition to select which paths are worthy of further exploration.

While this collaborative scaffolding proved effective, a natural question remains: can this process be conducted purely by AI? Specifically, if the necessary context is provided, does the AI possess the judgment to navigate \tradeoffmitigation chains as experts do? Prior work shows that LLMs perform well in automating research tasks with well-defined goals~\cite{si2024can,swanson2025virtual}. However, open-ended design tasks pose unique challenges; they are interdisciplinary, ill-defined, and lack objective criteria for optimization.  Progress in these domains depends heavily on tacit knowledge and expert judgment~\cite{polanyi1966logic}, making fully automated solutions potentially ill-suited.
Indeed, many of our experts often started with under-specified ideas and then worked iteratively with \sys to refine them.
Though our scope focused on establishing \tradeoffmitigation as a collaborative scaffold, a valuable direction for future work would be to directly evaluate if this reasoning structure can improve AI-only performance on creativity tasks like design ideation, or if AI can be trained to mimic the expert process of engaging with and navigating complex design solution spaces to produce high quality results.

\subsection{Supporting Mental Simulation and the Challenge of Verification}
Participants in our study highlighted a critical need to anticipate how ideas might fail before investing resources in them. The \tradeoffmitigation chains in \sys served as a scaffold for this mental simulation, allowing users to surface potential risks and approximate design outcomes. These findings contribute to a growing body of research using LLMs to test and approximate design ideas, offering preliminary validation for early-stage concepts~\cite{liu2024toward,makatura2023can}. 

However, the utility of these generated validations
% simulations 
is constrained by the reliability of the underlying model. Because LLMs are prone to hallucinations, our participants sometimes expressed hesitation in accepting generated critiques. They noted that without integrated sources or citations, they need to manually verify the AI's claims. Future systems must address this by integrating citation mechanisms to enhance credibility and reduce the cognitive load of verification.

Furthermore, there is a limit to how far text-based validations can push an idea. We observed that some participants would stop exploring within the tool once they felt a need for physical testing or empirical data. Successful designs emerge through repeated physical trials and thought exercises that often span long periods of time. While our study demonstrates the value of \sys in a concentrated session, future work should examine how such tools support the "slow hunch" of innovation over extended periods, bridging the gap between digital simulation and physical reality.

\subsection{Scaffolding Breadth and the Need for Dynamic Organization and Generation}
As shown by our study results, supporting broad exploration presents a dual challenge: users must be able to manage a large volume of ideas, while simultaneously pivoting into orthogonal solution spaces to discover novel directions. Our findings demonstrate that while \sys successfully supported the former and manual pivoting through its multi-threaded structure, the use of AI-driven abstraction to facilitate pivoting into orthogonal solution spaces revealed limitations. Users used the multi-threaded structure to manage ideas by manually adding, pruning, and positioning branches. This enabled a form of adaptive organization where the structure matched the user's intent. However, the abstraction feature lacked this flexibility, leading to a divide in utility. Novices struggled to understand and utilize the fixed abstract concepts, while experts, though more capable of using them, also expressed a desire to adjust the generated abstractions to better fit their mental models for the ideas.

This suggests that the limitation of abstraction feature was not merely a difference in comprehension due to expertise~\cite{chi1981categorization}, but also a lack of steerability in aligning the AI's output with the user's intent. Consequently, while abstractions are useful for pivoting into orthogonal solution spaces, the static concepts hinder users from effectively using them for pivoting into new solution spaces. Future work should therefore move toward dynamic scaffolding, where the generation and application of abstractions are interactive and user-steerable. Rather than rigid categorization, systems could allow users to edit abstract labels before the AI uses them to seed new ideas. By transforming abstraction from a fixed output into a negotiable parameter, future tools can ensure that the organization of the solution space evolves in sync with the user's developing mental model.

% \subsection{Tracking and Revisiting Ideas to Support Broad Exploration}
% Ideation often requires not only generating and analyzing ideas but also keeping track of them. Participants described how \sys helped them externalize and revisit their thinking, making it easier to move across chains and compare alternatives. The multi-threaded structure allowed users to record ideas in a compact format and to re-engage with branches that might otherwise be forgotten in linear chat interfaces.

% Participants also envisioned combining \sys with complementary tools to better realize the potential of its multi-threaded structure. They described workflows where they used a browser for information foraging while using \sys to organize and revisit ideas, suggesting opportunities for interface designs that separate information gathering from structured reasoning. Several participants also noted potential benefits for collaborative ideation. Having individual \tradeoffmitigation chains accessible before meetings could provide a shared grounding for discussion, and comparing different people’s chains could enable teams to explore the design space more broadly during the fuzzy front end of innovation~\cite{koen2001providing,reid2004fuzzy}. These observations suggest that structured multi-thread tracking may be valuable not only for individual ideation but also for coordination and shared understanding in team settings.

\subsection{Limitations}
Our study design and findings have several limitations. First, to balance data collection with user fatigue, the user study was limited to 30 minutes per task in both the controlled study and the expert condition. When using \sys in both settings, participants managed to explore many ideas but mentioned not having enough time to explore all the ideas provided by the system. Future work should examine how a longer time frame might affect participants’ ideation strategies and efficiency over time. Second, while we aimed to test whether \sys could support experts in their real-world ideation tasks, it was difficult to control for the varying difficulty of the tasks they proposed. This made it challenging to compare results across expert tasks. For this reason, we did not compare their behaviors with existing AI tools, as we did in the controlled study. A more rigorous study that directly compares the two with specific measures would provide greater insight into the value of \sys. Third, our participants consisted only of designers and engineers, but ideation is a common process across many fields of problem-solving. Future work could explore how \sys might assist people working in different domains.

\section{Conclusion}
Effective ideation requires not just generating ideas but articulating trade-offs, developing mitigations, and exploring multiple directions in parallel. We introduced \sys, an AI-augmented system that scaffolds multi-threaded \tradeoffmitigation chains to support this deeper and broader form of reasoning. By surfacing trade-offs, offering actionable mitigations, and organizing diverse idea threads within a structured workspace, \sys helps users refine ideas while continuing to expand the design space. A controlled study and expert case study show that \sys enables deep idea evaluation, actionable refinement, and broad exploration. These findings highlight the value of pairing LLMs with explicit reasoning structures that keep users actively engaged, rather than relying on directive automation.More broadly, our work points toward human–AI co-creation systems that support not only idea generation but the reflective, multi-threaded reasoning required for open-ended design, engineering, and R\&D tasks. Future work can extend this approach to longer-term workflows and collaborative settings.

\section{Acknowledgments}
This work was supported by the Toyota Research Institute, and partially with support from Google.org and the Google Cloud Research Credits program for the Gemini Academic Program.

\bibliographystyle{ACM-Reference-Format}
\bibliography{sample-base}

% \appendix
\section{Appendix}
\lstset{
  basicstyle=\ttfamily\small,
  breaklines=true,        % wrap long lines
  breakatwhitespace=true, % only break at spaces
  frame=single,
  columns=fullflexible,
  keepspaces=true
}

\subsection{Generative AI Usage}

We used ChatGPT to 1) generate the descriptions for the images, 2) for polishing the quality of text, and 3) formatting the tables.

\subsection{Design task description used in user study}

\subsubsection{Pre-defined condition}
\label{apd_task_description}
See table~\ref{table:design_tasks_predefine}.

\begin{table}[h]
\centering
\small
\renewcommand{\arraystretch}{1.3}
\begin{tabularx}{\linewidth}{cX}
\hline
\textbf{ID} & \textbf{Design Task in Pre-defined Condition} \\
\hline
1 & Clean laundry with less water. A standard washer can take up to 45 gallons per load, so an American household running 8–10 loads a week may use nearly 500 gallons—an unsustainable demand in water-scarce regions worldwide. Clothes need to be laundered because of smells and stains, but sometimes a full soak and cycle in a laundry machine is overkill for what needs to be done. We're looking for ways to "refresh" clothes in order to keep them ready to wear. How would you get clothes cleaned up with less water? \\
2 & Minimize accidents from people walking and texting on a cell phone. Distracted walking can be life-threatening. People texting and walking were four times more likely to display “unsafe walking behavior” than other pedestrians. Walking while texting caused more than 11,000 injuries in 2019 and led to over 5,000 deaths. Although most people recognize the risk, many continue to text while walking, often without realizing it. How would you help minimize accidents caused by walking and texting? \\
\hline
\end{tabularx}
\caption{Design tasks used in pre-defined condition}
\label{table:design_tasks_predefine}
\end{table}

% \FloatBarrier 
% \subsubsection{Expert condition}
% See table~\ref{table:design_tasks_expert}
% \label{apd_expert_task_des}
% \begin{table}[h]
% \centering
% \small
% \renewcommand{\arraystretch}{1.3}
% \begin{tabularx}{\linewidth}{cX}
% \hline
% \textbf{ID} & \textbf{Design Task in Expert Condition} \\
% \hline
% 1 & Tools that compensate for hand tremors in fine work. \\
% 2 & When you’re at sea at night, it’s important to recognize navigation lights of other ships. This is difficult to practice except at sea because the difficulty comes from the movement of the boats in the waves, fog/haze, the relative direction of the boats, and everything is in motion. Also, you have to memorize the light configurations (tug, fishing boat, etc.). How to learn and practice this? \\
% 3 & Design a way for autonomous cars to reliably transfer perception data to each other as they drive in an environment with human passengers. \\
% 4 & Golf cart that drives itself and follows the golfer naturally. \\
% 5 & Recognizing hand gestures from continuous hand interaction in live presentation settings. \\
% 6 & Design a system that automatically adjusts a car seat to the driver’s most comfortable position, eliminating the need for manual adjustment. \\
% \hline
% \end{tabularx}
% \caption{Design tasks used in expert condition}
% \label{table:design_tasks_expert}
% \end{table}

\FloatBarrier 
\subsection{Baseline Annotation CodeBook}
\label{apd_codebook}
\begin{table}[t]
\centering
\footnotesize
\begin{tabularx}{\textwidth}{@{} l l X X @{}}
\toprule
\textbf{Type} & \textbf{Category} & \textbf{Definition} & \textbf{Example} \\
\midrule
action & direct idea generation using ChatGPT or search & User explicitly asks ChatGPT or a search tool for new ideas. & P16 prompts ChatGPT: "How would you help minimize accidents caused by walking and texting?" \\ \midrule
action & add users own idea & User introduces a novel idea without ChatGPT prompting. Not merely adopting someone else's idea. & P8 adds their own idea: "Walking alert — Notifications/alarms based on motion sensitivity to make user aware of their distracted walking." \\ \midrule
action & analyze tradeoff using ChatGPT or search & User asks ChatGPT/search to weigh pros/cons or consequences. & P14 prompts ChatGPT: "Come up with possible user pain points for target cleaning." \\ \midrule
action & analyze tradeoff themselves & User personally evaluates pros/cons without asking ChatGPT or search. & P8 thinks out loud: "AR Obstacle Hunt, but people will focus on using AR, end up using phones even more." \\ \midrule
action & find solution to certain tradeoff using ChatGPT or search & User asks ChatGPT/search for ways to overcome a limitation or drawback. & P3 prompts: "Is it possible to design different cleaning load for different cloth amount?" \\ \midrule
action & find solution to certain tradeoff themselves & User generates their own mitigation to a limitation or drawback. & P19 addresses the tradeoff "blocking social media would be irritating for users": "Instead of simply blocking social-media applications, we can also send vibrational feedback to the user's phone to remind them about their motion." \\ \midrule
action & find similar idea using ChatGPT or search & User prompts ChatGPT/search to provide similar or variant ideas. & P6 prompts ChatGPT: "Can you think of more ideas similar to the first one?" \\ \midrule
action & find similar idea themselves & User generates their own variants of an idea without asking ChatGPT/search. & P4 writes: "Red light audio to remind pedestrians, like 'wait'," after viewing the idea "Adaptive Haptic Landmarks." \\ \midrule
action & ask question & User seeks general information, facts, or explanations (not necessarily ideas). & P16 prompts: "What sensors are in a washing machine?" \\ \midrule
action & other & Other actions not belonging to the above categories. & P12 thinks out loud: "This question should be reframed as 'How might we clean/sanitize lightly soiled clothes using as little water as possible'." \\ \midrule
information & ideas & Concrete solutions, design options, or concepts (from ChatGPT/search or user notes). Split multi-idea responses into separate items. & "Targeted Spot and Steam Module." \\ \midrule
information & tradeoffs & Limitations, possible failures, or conflicts noted by ChatGPT/search or the user. & P15 thinks out loud: "Energy vs water tradeoff: some water-free methods (steam, ultrasonic) use significant electricity." \\ \midrule
information & other knowledge & Background information, facts, or context not fitting ideas/tradeoffs. & ChatGPT answers: "Modern washing machines use sensors such as water level, water temperature, flow, load/imbalance, vibration, motor speed, turbidity/soil, door lock, leak, pH/conductivity, and humidity/moisture sensors." \\ \midrule
links & action to information, or information to action & Relationships between action and information using two criteria: whether the actions addressed the same design idea, and whether one action contributed to developing the idea, tradeoff, or knowledge. & — \\
\bottomrule
\end{tabularx}
\caption{Codebook for the baseline annotation, including taxonomy examples across actions, information, and links.}
\label{tab:taxonomy_examples_no_participants}
\end{table}

\FloatBarrier 
\subsection{Prompt}
\FloatBarrier 
\subsubsection{Prompt 1} Tradeoff Generation
\label{prompt_tradeoff_generation}
\begin{lstlisting}
You are a design expert evaluating a proposed mechanism for a specific design problem.
Your task: Identify mechanism-specific trade-offs that limit the effectiveness or feasibility of this solution.

Instructions:
- List the distinct trade-offs. These are specific downsides or limitations that arise because of this particular mechanism, not general design challenges or unrelated drawbacks.
- Avoid overlapping or redundant trade-offs. Each trade-off should address a different type of concern
- For each trade-off, provide:
    - A short, clear name that captures the essence of the issue.
    - A brief description explaining the negative impact and why it matters within 50 words.

Input:
design problem: {design_problem}
mechanism: {mechanism}

Output:
- Give all the main tradeoffs you found followling the instruction.
- Choose the top three among the listed tradeoffs. For those three, Your output should use this table format and provide your table between the tags :  < table > The tradeoff table </ table >:
    Here's an example:
    < table >
    | id | name       | description |  
    |----|---------------|-------------|  
    | 1  | [tradeoff Name]  | [Impact 1]  |  
    | 2  | [tradeoff Name]  | [Impact 2]  |  
    | 3  | [tradeoff Name]  | [Impact 3]  |  
    </ table >
- do not add symbols like '**' in the table
\end{lstlisting}
\FloatBarrier 

\subsubsection{Prompt 2} Solution Generation
\label{prompt_solution_generation}
\begin{lstlisting}
You are a design expert analyzing a given mechanism for a specified design problem.
\*\* task \*\*: Give solutions to solve the trade-off listed below. You can add new functions to the original mechanism or give a total different mechanism that would address the constraints.
\*\* requirement \*\*:
\- **Uniqueness and creativity check** : Each solution should be creative and substantially different from every *other* solution you provide.

```
- **Feasibility filter** :
    - Assume current or near-future technology.  
    - If feasibility is questionable, revise or replace the idea.

- **Trade-off focus** :
    - Make it clear *how* the proposal directly addresses the stated trade-off.  
    - You can make incremental change of current mechanism or give a total different mechanism.
```

* \*\* Proper level of detail \*\*: Explain the mechanism clearly but not exhaustively, leaving space for subsequent ideation while still providing guidance.

```
- For each solution, give a concise name and a brief description of the idea within 50 words.
```

Input:
design problem: {design\_problem}
mechanism: {mechanism}
trade-off: {tradeoff}

Output:

* Give all the top solutions you found followling the instruction.
* Choose the top three among the listed solutions. For those three, Your output should use this table format and provide your table between the tags :  < table > The solution table \</ table >:
  Here's an example:
  < table >

  | id          | name             | description   |
  | ----------- | ---------------- | ------------- |
  | 1           | \[solution Name] | \[solution 1] |
  | 2           | \[solution Name] | \[solution 2] |
  | 3           | \[solution Name] | \[solution 3] |
  | \</ table > |                  |               |
* do not add symbols like '\*\*' for the texts in the table
  \end{lstlisting}

\FloatBarrier

\subsubsection{Prompt 3} Abstraction Generation
\label{prompt_abstract_generation}
\begin{lstlisting}
You are an assistant good at abstracting concepts and inferring user intent.
The user is working on the following design problem: {design\_problem}
They have identified the following mechanism of interest: {mechanism}

## Task:

```
Infer the most relevant high-level concepts that reflect what the user is actually interested in for solving the given problem.
e.g. for the idea of using lemon spray to solve 'clean launry with less water', the concept of lemon spray could be 'target the stain', 'use natural ingredients', 'mask the smell'
```

## Requirements:

1. Each concept should be different from the others.
2. The concept should have a level of abstarction but avoid overly broad or repeating the original idea or the problem.
3. Consider all the possible concepts and choose the top-{concept\_num} concepts that are most relevant to the given mechanism and most useful for solving the problem.

## Format

```
1.Output your findings as a markdown table
- The table should have the following columns:
    - **name**: A concise, descriptive label for the concept.
    - **description**: A more detailed explanation of the concept.
2.only top-{concept_num} concepts.
```

Your output should use this table format and provide your table between the tags :  < table > The concept table \</ table >:
Here's an example:
< table >
\| name | description |
\|--------|------------|
\| Your first concept  | A concise description of the concept |
\| Your second concept | A concise description of the concept |
\</ table >

* do not add symbols like '\*\*' for the names in the table
  \end{lstlisting}

\FloatBarrier 

\subsubsection{Prompt 4} Relevant Abstraction Retrieval
\label{prompt_abstract_retri}
\begin{lstlisting}
You are an assistant good at abstracting concepts and inferring user intent.
The user is working on the following design problem: {design\_problem}
They have identified the following mechanism of interest: {mechanism}

## Task:

```
Infer the most relevant high-level concepts that reflect the high-level concepts the user is actually interested in for solving the given design problem. The concept must be the categories in the provided list.
```

## Mechanism list: {mechanism\_list}

## Format:

```
1.Output your findings as a markdown table
- The table should have the following columns:
    - **concept_name**: The exact concept name from the list.
    - **reason**: A brief reason for why choosing the concepts.
```

Your output should use this table format and provide your table between the tags :  < table > The tradeoff table \</ table >:
Here's an example:
< table >
\| name        | reason                                      |
\|---------------------|----------------------------------------------------------|
\| Your first concept  | A brief reason              |
\| Your second concept | A brief reason             |
\</ table >

## Requirements:

```
1. The selected concepts must be highly related to the given mechanism.
2. Use the exact name from the provided list.
3. If you can not find any relevant concept in the list, output one sentence 'No concept found' instead of a table.
```

\end{lstlisting}

\FloatBarrier 
\subsubsection{Prompt 5} Abstraction Redundant Check
\label{prompt_abstract_check}
\begin{lstlisting}
For the two given lists of concepts, there are all for solving the problem: {design\_problem}.
Identify items from List 1 that has the same meaning as any item in List 2 with similar expression.

## Input:

List 1: {new\_list}
List 2: {original\_list}

## Output:

For every match, return a single row in the table below, wrapped in the tags shown.
Use the *exact* mechanism name from each list---do **not** paraphrase. here's an example:
< table >
\| name1 | name2 |
\|--------|--------|
\| the first matched concept name in list1  | the relevant concept name in list2 |
\</ table >

## Constraints:

Only output the matched concepts. If no similar items are found, output nothing.
Ensure the output strictly follows the required format.
\end{lstlisting}

\FloatBarrier 

\subsubsection{Prompt 6} Similar Idea Generation
\label{prompt_similar_gen}
\begin{lstlisting}
You are a design expert working on a design problem: {design\_problem}.
Think of more specific sub mechanisms for the given high-level mechanism: {mechanism}

## Requirements

```
1. The submechanisms should demonstrate the given abstract mechanism well with more concrete ideas for solving the given problem.
2. The sub mechanisms should be craetive and different from each other.
3. Generate {mech_num} sub mechanisms.
4. Proper level of detail: Explain the mechanism clearly but not exhaustively, leaving space for subsequent ideation while still providing guidance.
```

## Format

```
1.Output your findings as a markdown table
- The table should have the following columns:
    - **name**: A concise, descriptive label for the submechanism.
    - **description**: A more detailed explanation of the submechanism.
Your output should use this table format and provide your table between the tags :  < table > The sub-mechanism table </ table >:
Here's an example:
< table >
    |  name        | description                                      |
    |---------------------|----------------------------------------------------------|
    | Your first sub mechanism  | A concise description              |
    | Your second sub mechanism | A concise description            |
</ table >
2.do not add symbols like '**' for the texts in the table
```

\end{lstlisting}

\FloatBarrier 

\subsubsection{Prompt 7} Answer Generation
\label{prompt_answer_gen}
\begin{lstlisting}
You are a design expert working on a design problem: {design\_problem}.
The most recent content you are working on is: {idea}
Consider the context given, answer the question relevant to the design problem: {question}

## Requirements:

The answer should be concise and to the point, and should be no more than 50 words.

## Format:

Provide your answer between the tags: < answer > and \</ answer >
Here's an example:
< answer > \[answer] \</ answer >
\end{lstlisting}

% \begin{figure}[t]
%   \centering
%   \includegraphics[width=\linewidth]{parts/figure/system_trees.png}
%   \caption{Participants’ ideation process in \sys represented as a tree structure in the controlled user study}
%   \label{fig:system_tree}
% \end{figure}

% \begin{table}[!ht]
%     \centering
%     \begin{tabular}{llllll}
%     \hline
%         ~ & results worth effort. & expressiveness & exploration & Immersion & enjoyment  \\ \hline
%         Average & 9.67 & 8.83 & 9.17 & 9.17 & 10.00  \\ 
%         SD & 0.516 & 1.329 & 1.329 & 1.329 & 0.000 \\ \hline
%     \end{tabular}
%     \caption{Creeativity Support Index in Expert condition}
%     \label{expert_CSI}
% \end{table}

% \begin{wrapfigure}{r}{0.55\linewidth}
%   \centering
%   \includegraphics[width=\linewidth]{parts/figure/expert_process_rating.png}
%   \caption{Average ratings of participants' evaluation of the ideation process using \sys in the expert condition.}
%   \label{fig:survey_process}
% \end{wrapfigure}

% Preamble:
% \usepackage{booktabs}
% \usepackage{tabularx}

\FloatBarrier

\subsubsection{Prompt 8} Schema Generation
\label{prompt_schema_gen}
\begin{lstlisting}
You are a creative design expert known for rapid, divergent ideation and adopting inspirations from diverse domains.

Task:Generate exactly 10 high-level solution directions (broad categories) for the given design challenge.

Challenge:{design_problem}

Thinking Guidelines

Examine the problem from multiple lenses.

Optionally adopt inspirations from other domains to spark fresh directions.

Make each direction conceptually distinct; avoid overlap.

Keep each description concise (<= 30 words).

Required OutputReturn a pure JSON array (no markdown, no commentary) with 10 objects.Each object must use the following schema:
{
  "direction": "<concise category label>",
  "description": "<explanation of how this direction addresses the task with 30 words>"
}
\end{lstlisting}
\FloatBarrier 

\subsubsection{Prompt 9} Schema Check
\label{prompt_schema_check}
\begin{lstlisting}
You are given an initial list of high-level solution directions for the design problem.

Input:{directions_output}

Task: Refine this list so the categories are mutually distinct, collectively cover a wide range of approaches, and remain high-level (i.e., they describe opportunity spaces, not specific implementations).

Instructions

Evaluate each existing category:

Is it conceptually different from the others?

Is the description broad enough (no single technology or tactic baked in)?

Identify overlaps, gaps, or narrow definitions.

Produce an improved list of 10 categories that:

Span diverse directions.

Are mutually exclusive yet collectively exhaustive.

Invite multiple implementation ideas.

For every revised category, provide:
1.name (2-5 words)
2.description (1-2 sentences; keep implementation-agnostic)

Output Format
{
  "categories": [
    {
      "name": "",
      "description": ""
    }
  ]
}
\end{lstlisting}

\subsubsection{Prompt 10} Idea Generation
\label{prompt_schema_idea_gen}
\begin{lstlisting}
You are a seasoned creative-design expert renowned for rapid, divergent ideation.

Context: You now have a refined set of high-level categories describing opportunity spaces for the design problem. Your task is to populate these categories with concrete solution ideas.

Design Problem:{design_problem}

Input:{categories_output}

Guidelines for Each Category:

Quantity: Provide exactly five distinct solution ideas.

Inspiration: You may adopt inspirations from other domains, but keep solutions in the original problem context. If outside inspiration is used, mention it briefly in parentheses.

Level of Detail: Write 20--40 words per idea---concrete enough to convey the mechanism, but still leaving room for later elaboration.

Diversity Check: Ensure the five ideas within the same category are substantially different from one another.

Output Format:
[
  {
    "id": "1",
    "name": "<category name>",
    "description": "<category description>",
    "mechanisms": [
      {
        "name": "<concise idea label>",
        "description": "<20-40-word description. Optional outside inspiration in parentheses>"
      },
      {
        "name": "...",
        "description": "..."
      }
    ]
  }
  /* repeat for each category */
]
\end{lstlisting}

\FloatBarrier 
\subsection{User study Results}
\subsubsection{Tree visualization for P11 to P20} See Figure~\ref{fig:all_tree_2}.
\label{all_tree_2}
\begin{figure}[t]
  \centering
  \includegraphics[width=\linewidth]{parts/figure/all_trees_2.png}
  \caption{10 Participants’ (P11-P20) ideation processes in the controlled user study are represented as tree structures. See the reset of the data in the Appendix. Blue dots represent solutions, orange dots represent tradeoffs, and yellow dots represent other information. In the baseline condition, grey dots represent the initial prompt or search queries used to start the tree. These are not counted as tree nodes but are included to illustrate how the trees are constructed.}
  \Description{Two rows of small node-and-branch “trees,” one labeled “FlexMind” (top) and one labeled “Baseline” (bottom). Each row shows ten columns labeled p1–p10. Within each participant column are one or more compact trees made of colored dots connected in simple branches. Blue dots mark solution nodes, orange dots mark trade-off nodes, and yellow dots mark other information (such as questions or similar-by-schema ideas). In the Baseline row, light gray dots appear at the tops of some trees to indicate the initial prompt or search query; these gray dots are not counted as tree nodes. Compared to Baseline, the FlexMind row shows more and taller trees with more branching and more orange nodes.}
  \label{fig:all_tree_2}
\end{figure}

\subsubsection{ICC for expert rating}
See table~\ref{table:expert_rating_icc}.

\label{ICC}
\begin{table}[h]
\centering
\small
\renewcommand{\arraystretch}{1.2}
\setlength{\tabcolsep}{6pt}
\begin{tabular}{lcc}
\hline\hline
 & Task 1 (n=112) & Task 2 (n=161) \\
\hline
Novelty     & 0.7701 & 0.7658 \\
Feasibility & 0.7626 & 0.7896 \\
Value       & 0.5558 & 0.7037 \\
\hline\hline
\end{tabular}
\caption{Inter-rater reliability (ICC(2,k)) for expert ratings of novelty, feasibility, and value across the two tasks. $n$ indicates the number of ideas rated per task.}
\label{table:expert_rating_icc}
\end{table}
\end{document}